\documentclass[a4paper,fleqn,usenatbib]{mnras}

\usepackage{newtxtext,newtxmath}

\usepackage[T1]{fontenc}
\usepackage{ae,aecompl}

\usepackage{graphicx}	
\usepackage{amsmath}	
\usepackage{amssymb}	

\title[Swift/XRT- NuSTAR spectra of type 1 AGN]{Swift/XRT-  NuSTAR spectra of type 1 AGN: confirming INTEGRAL results on the high energy cut-off}

\author[M. Molina et al.]{
M. Molina,$^{1}$\thanks{E-mail: manuela.molina@inaf.it}
A. Malizia,$^{1}$
L. Bassani,$^{1}$
F. Ursini,$^{1}$
A. Bazzano,$^{2}$
P. Ubertini$^{2}$
\\
$^{1}$INAF-OAS Bologna, via Gobetti 101, I-40129 Bologna, Italy\\
$^{2}$INAF-IAPS Roma, Via Fosso del Cavaliere 100, I-00133 Roma, Italy\\
}

\date{Accepted 2019 January 8. Received 2019 January 8; in original form 2018 October 24}

\pubyear{2018}

\begin{document}
\label{firstpage}
\pagerange{\pageref{firstpage}--\pageref{lastpage}}
\maketitle

\begin{abstract}
We present the 0.5 -- 78 keV spectral analysis of 18 broad line AGN belonging to the  {\it INTEGRAL} complete sample. Using simultaneous 
{\it Swift}-XRT and  {\it NuSTAR} observations and employing a simple phenomenological model to fit the data, we measure with a 
good constraint the high energy cut-off in 13 sources, while we place lower limits on 5 objects. We found a mean high-energy cut-off of 111 keV ($\sigma$ = 45 keV) for the whole sample,
in perfect agreement with what found in our previous work using non simultaneous observations and with what recently published using {\it NuSTAR} data.
This work suggests that simultaneity of the observations in the soft and hard X-ray band is important but not always essential, especially if flux and spectral variability are properly accounted for. 
A lesser agreement is found when we compare our cut-off measurements with the ones obtained by Ricci et al. (2017) using {\it Swift}-BAT high energy 
data, finding that their values are systematically higher than ours.
We have investigated whether a linear correlation exists between photon index and the cut-off and found a weak one, probably to be ascribed to the non perfect modelling of the soft part of the 
spectra, due to the poor statistical quality of 
the 2-10 keV X-ray data. No correlation is also found between the Eddington ratio and the cut-off, suggesting that only using high statistical quality broad-band spectra  is it possible to verify the 
theoretical predictions and study the physical characteristics of the hot corona and its geometry.

\end{abstract}




\section{Introduction}
In recent years, high energy observations of Active Galactic Nuclei (AGN) have 
provided new insights on the physics and mechanisms at play in some of the most 
energetic phenomena in the Universe. In particular, the slope of the continuum emission and its high energy cut-off are essential 
for spectral modelling of AGN, since these parameters directly probe the 
physical characteristics of the Comptonizing region around the central nucleus.
Indeed the X-ray continuum of AGN can be ascribed to the inverse Compton 
scattering of soft photons arising from the accretion disc by energetic 
electrons thermally distributed above the disc, the so-called X-ray corona. 
Clearly to have an overview of the physics and structure of the corona we 
need to study the broad-band spectra of a large sample of AGN in order to 
account for all spectral components, remove the degeneracy between parameters 
and therefore being able to obtain a precise estimate of the photon index and high energy cut-off 
for a large number of objects. 

Many independent observations in the X-rays suggest that the corona should be compact and 
located very close to the black hole, but its characteristics are  still largely unknown as only recently 
information on the electron temperature (cut-off energy) and the optical depth 
(photon index) of the hot plasma in the corona has become available.
Furthermore, very little is known about the heating and thermalization mechanisms operating in the corona, i.e. 
what controls the plasma temperature and how any stable equilibrium is reached 
(\citealt{Fabian_2015}, \citealt{Fabian_2017}). 
Recently,  \citet{Malizia:2014} and \citet{Ricci_2017}
have analysed large samples of type 1 AGN and, through
broad-band spectra covering a large energy range (typically from few keV up to hundreds of keV), found 
that the cut-off energy distribution peaks at around 100\,keV, a result 
which is consistent with synthesis models of the Cosmic X-ray Background
which locate the cut-off below 200 keV in order not to exceed it \citep{Gilli:2007}.
These first estimates of  the temperature of the corona and its compactness parameter\footnote{This is a dimensionless 
parameter defined as l=(L/R)*($\sigma_T$/m$_{\rm e}$c$^3$), 
where L is the source luminosity, R its radius, $\sigma_T$ the Thomson cross section 
and m$_{\rm e}$ the mass of the electron} indicate that this region is hot and radiatively compact, 
that pair production and annihilation are essential
mechanisms operating in it and that thermal and non thermal particles are present (\citealt{Fabian_2015}; \citealt{Fabian_2017}). 

\begin{table*}
\small
\begin{center}
\centerline{{\bf Table 1 - Observation Log}}
\begin{tabular}{lcccc}
\hline
{\bf Source}    & {\bf XRT Obs. Date}  &  {\bf XRT exposure }  &  {\bf NuSTAR Obs. Date}  &  {\bf NuSTAR exposure }\\
                &                      &          (ksec)       &                          & (ksec)    \\
\hline
QSO B0241+62     &  30 July 2016       &     6.5                                 & 31 July 2016                     & 23.3 \\
MCG+08-11-011    &  18 Aug. 2016       &    15.9                                 & 18 Aug. 2016                    & 97.9  \\
Mrk 6            &  08-10 Nov. 2015    &     15.8                                  & 9 Nov. 2015                     & 43.8 \\
FRL 1146         & 28 July 2014        &      6.1                                     &27 July 2014                &    21.3     \\
IGR J12415-5750  &  27 Apr. 2016       &      5.7                                   & 27 Apr. 2016                    & 16.4\\      
4U 1344-60       & 17 Sep. 2016        &       1.8                                  & 17 Sep. 2016                    &99.5 \\
IC 4329A         & 14 Aug. 2012        &        2.2                                 &12  Aug. 2012                    & 16.2\\
IGR J16119-6036  & 27 Apr. 2018        &       6.8                                  &28 Apr. 2018                     &  22 \\
IGR J16482-3036  & 3 Apr. 2017         &        7.1                                  & 3 Apr. 2017                       & 18.9\\
GRS 1734-292     & 16 Sep. 2014        &     6.2                                   & 16 Sep. 2014                  &20.2\\
2E 1739.1-1210   & 17 Oct. 2017        &     6.9                                   & 17 Oct. 2017                    &21.4 \\
IGR J18027-1455  & 2 May 2016          &     6.6                                    &2 May 2016                      &19.9\\
3C 390.3         & 25 May 2013         &    2.1                                      & 24 May 2013                   &47.6\\  
2E 1853.7+1534   & 13 Oct. 2017        &   6.4                                       &13 Oct. 2017                     &21.3 \\    
NGC 6814         &05 July 2016         &   7.6                                       & 04 July 2016                    &148.4 \\
4C 74.26         & 21 Sep. 2014        &    2.2                                      & 21 Sep. 2014                    & 19.0\\  
                 & 24 Sep. 2014        &     2.0                                     & 22 Sep. 2014                     &56.5 \\
                 & 31 Oct. 2014        &       1.8                                   & 30 Oct. 2014                     &90.9\\
                 & 22 Dec. 2014        &        2.1                                 & 22 Dec. 2014                     &42.7\\
S5 2116+81       & 5 Aug. 2017         &       6.7                                  &     5 Aug. 2017               & 18.5\\             
IGR J21247+5058  & 13 Dec. 2014        &       6.8                                 &13 Dec. 2014               &  24.3\\
\hline
\end{tabular}

\end{center}
\label{log}
\end{table*}

However, these initial  works made use of non-simultaneous low versus high energy data, and despite the introduction of cross-calibration
constants to properly take into account flux variability (and possible mismatches in the instruments calibration),
some degree of uncertainty remains due to the fact that the broad-band
spectra are a combination of snapshot observations in the soft X-rays (typically from {\it XMM-Newton} or from {\it Swift}-XRT) with
time-averaged (on timescales of years) spectra at high energies ({\it INTEGRAL}-IBIS, {\it Swift}-BAT). 
Despite the non-simultaneity of the data, cross-calibration constants are typically found to be close to 1 
(\citealt{Molina:2009}; \citealt{ Malizia:2014}), 
hinting at the fact that long-term flux variability can be easily accounted for in modelling of 
spectra taken in different epochs. Spectral variability, however, cannot be excluded {\it a priori}, making this the 
only uncertainty on an otherwise statistically robust result. If one wants to remove this 
ambiguity, it is fundamental to have simultaneous observations in both the soft and hard 
X-ray bands. This is now achievable thanks to the advent of {\it NuSTAR}, which is complementary,
at least for bright sources, to {\it INTEGRAL}/IBIS and {\it Swift}/BAT in combination with soft X-ray data.
In fact, several studies have been published recently which confirm,
through simultaneous, high quality broad-band {\it NuSTAR}/{\it XMM-Newton} observations, what was previously found
using non-contemporaneous data.
However, it is not easy to obtain this type of high quality broad-band spectra and it is for this reason
that sometime non-contemporaneous {\it NuSTAR}/soft X-ray  observations are
used preferring quality over simultaneity (e.g. \citealt{Tortosa:2017}).
 
The main objective of this paper is to verify what was previously established from the broad-band
spectral analysis of non-simultaneous {\it XMM-Newton}, {\it INTEGRAL}/IBIS and {\it Swift}/BAT
spectra, in order to establish whether variability has still some effect on the spectral fits. 
If we find that for a significant sample of AGN the cut-off values are similar, either using simultaneous broad-band observations or not,
than we can safely assume that spectral variability is not a big issue.
Also, we aim at putting firmer constraints on the value of the high energy 
cut-off. Besides, we also attempt to establish whether
choosing to use relatively simple, phenomenological models  (like the
\texttt{pexrav} model, which is more suitable for low statistics broad-band 
spectra) over more complex and more physical ones can nonetheless provide 
accurate measurements of the high energy cut-off.

\begin{table*}
\begin{center}
\vspace{0.2cm}
\begin{tabular}{lccccccccc}
\multicolumn{10}{c}{{\bf Table 2- \texttt{pexrav} model fit results}} \\
\hline
{\bf Name}      & {\bf N$_{\rm \bf H}^{\rm \bf FC}$} &{\bf N$_{\rm \bf H}^{\rm \bf PC}$ (cf)} & {\bf $\Gamma$}    &{\bf E$_{\rm \bf c}$}  &{\bf R}&  {\bf EW}& {\bf C$_{\rm \bf FPMA}$}        &    {\bf C$_{\rm \bf FPMB}$ } & {\bf $\chi^{2}_{\nu}$}    \\
                           &     (10$^{22}$ cm$^{-2}$)                      & (10$^{22}$ cm$^{-2}$)                     &                              &(keV)                        &           &         (eV)&                              &                             &                 \\
\hline
QSO B0241+62    & 0.32$^{+0.23}_{-0.19}$  &             --             & 1.82$^{+0.07}_{-0.06}$ & $>$ 198                & 0.72$^{+0.35}_{-0.28}$   & 85$^{+30}_{-28}$&  1.53$^{+0.21}_{-0.15}$ & 1.56$^{+0.22}_{-0.18}$ & 1.00 \\
MCG+08-11-011$^{\star}$&    --                &                 --       & 1.80$^{+0.01}_{-0.01}$ &  163$^{+53}_{-32}$ & 0.36$^{+0.09}_{-0.08}$& 128$^{+17}_{-17}$&1.19$^{+0.03}_{-0.03}$  &  1.24$^{+0.03}_{-0.03}$  &  1.01 \\
Mrk 6           &             --          & 9.92$^{+1.73}_{-7.01}$  (0.95$^{+0.02}_{-0.02}$) & 1.73 (fixed)   & 120$^{+51}_{-28}$  & 1.33$^{+0.41}_{-0.37}$ & 82$^{+31}_{-30}$&1.44$^{+0.13}_{-0.11}$   & 1.47$^{+0.13}_{-0.11}$ & 0.93\\
                &                                                 & 22.4$^{+8.60}_{-6.70}$  (0.51$^{+0.05}_{-0.05}$) &                      &                                 &                                      &             &                           & & \\
FRL 1146        &  0.70$^{+0.08}_{-0.13}$ &               --                                                        &  2.07$^{+0.09}_{-0.07}$  &$ >$182 & 1.03$^{+0.42}_{-0.32}$ & 144$^{+44}_{-44}$&1.09$^{+0.09}_{-0.08}$ & 1.14$^{+0.09}_{-0.09}$ & 1.07 \\ 
IGR J12415-5750 &          --             &               --                                                              & 1.63$^{+0.05}_{-0.05}$  & 123$^{+54}_{-47}$  & $<$0.23                        & 65$^{+20}_{-30}$&1.41$^{+0.09}_{-0.07}$          & 1.45$^{+0.09}_{-0.08}$   & 0.97 \\
4U 1344-60$^{\star}$    &           --           &  1.68$^{+0.59}_{-0.22}$ ($>$0.83)                       & 1.91$^{+0.05}_{-0.03}$  & 141$^{+46}_{-26}$   & 1.16$^{+0.19}_{-0.16}$ &  109$^{+21}_{-19}$& 1.13$^{+0.10}_{-0.08}$& 1.19$^{+0.10}_{-0.09}$ &1.08\\
IC 4329A$^{\star}$      & 0.54$^{+0.06}_{-0.05}$&          --                                              &  1.73$^{+0.01}_{-0.01}$     &  153$^{+20}_{-16}$  &  0.43$^{+0.05}_{-0.05}$ & 83$^{+9}_{-8}$& 1.17$^{+0.06}_{-0.05}$&  1.22$^{+0.06}_{-0.06}$ & 1.01\\
IGR J16119-6036 &           --            &          --                                                                &1.57$^{+0.08}_{-0.04}$     & 69$^{+102}_{-26}$  & 0.40$^{+0.56}_{-0.34}$ &147$^{+52}_{-49}$ &0.95$^{+0.07}_{-0.06}$ &0.96$^{+0.07}_{-0.06}$ &1.09\\
IGR J16482-3036 &           --            &             --                                                                & 1.63$^{+0.11}_{-0.18}$     &  $>$38                 &             0 (fixed)                          &163$^{+173}_{-136}$&1.53$^{+0.34}_{-0.28}$   &1.59 $^{+0.34}_{-0.29}$      & 0.87\\
GRS 1734-292    & 1.08$^{+0.15}_{-0.18}$  &            --                                                                  & 1.64$^{+0.06}_{-0.06}$       &   53$^{+13}_{-9}$  &   0.45$^{+0.28}_{-0.24}$          & 34$^{+23}_{-24}$   &0.98$^{+0.05}_{-0.05}$       & 1.07$^{+0.05}_{-0.05}$    & 1.07\\
2E 1739.1-1210  & 0.16$^{+0.06}_{-0.05}$  &             --                                                             & 1.91$^{+0.09}_{-0.08}$      & $>$110              & 0.78$^{+0.65}_{-0.48}$         & 94$^{+46}_{-45}$&1.09$^{+0.08}_{-0.08}$  & 1.14$^{+0.09}_{-0.08}$ & 1.06 \\
IGR J18027-1455 &   --                    &  0.59$^{+0.26}_{-0.12}$ ($>$0.79)                     & 1.78$^{+0.07}_{-0.07}$ & $>$91           & $<$0.39                                 & 138$^{+34}_{-35}$&1.07$^{+0.06}_{-0.06}$ & 1.08$^{+0.06}_{-0.06}$ & 1.03 \\
3C 390.3$^{\star}$        & $<$0.82            &      --                                                                    & 1.65$^{+0.02}_{-0.02}$ & 130$^{+42}_{-32}$  &  $<0.14$                   & 77$^{+22}_{-20}$&1.15$^{+0.05}_{-0.05}$ & 1.20$^{+0.06}_{-0.05}$ & 0.99\\
2E 1853.7+1534  &   --                   &  0.58$^{+0.53}_{-0.17}$ ($>$0.73)                & 1.59$^{+0.18}_{-0.14}$     & 42$^{+22}_{-11}$   & 1.29$^{+1.31}_{-0.75}$     & 96$^{+57}_{-54}$&1.13$^{+0.11}_{-0.11}$ & 1.19$^{+0.12}_{-0.11}$ & 0.96 \\
NGC 6814        & 0.03$^{+0.02}_{-0.02}$ &        --                                                                  & 1.68$^{+0.02}_{-0.02}$     & 115$^{+26}_{-18}$ &  0.32$^{+0.09}_{-0.08}$  & 111$^{+10}_{-10}$&1.14$^{+0.04}_{-0.04}$ & 1.14$^{+0.04}_{-0.04}$ & 1.09\\
4C 74.26 (1)    & 0.11$^{+0.03}_{-0.03}$ &        --                                                                  & 1.80$^{+0.06}_{-0.05}$     & 94$^{+54}_{-26}$   & 0.63$^{+0.32}_{-0.27}$   & 60$^{+26}_{-25}$&1.14$^{+0.04}_{-0.04}$ & 1.15$^{+0.07}_{-0.07}$ & 0.88 \\
         (2)    &   --                   &        --                                                                  & 1.69$^{+0.03}_{-0.03}$     & 71$^{+12}_{-9}$     & 0.50$^{+0.13}_{-0.12}$   & 57$^{+15}_{-14}$&1.22$^{+0.07}_{-0.06}$  & 1.25$^{+0.07}_{-0.06}$ & 1.01 \\
         (3)    &   --                   &      --                                                                   & 1.72$^{+0.03}_{-0.03}$     & 115$^{+58}_{-29}$  & 0.23$^{+0.16}_{-0.09}$   & 63$^{+14}_{-12}$&1.26$^{+0.08}_{-0.07}$  & 1.28$^{+0.08}_{-0.07}$ & 1.02 \\
         (4)    & 0.11$^{+0.03}_{-0.03}$ &        --                                                                 & 1.75$^{+0.03}_{-0.03}$     & 119$^{+48}_{-27}$  & 0.10$^{+0.10}_{-0.12}$   & 52$^{+18}_{-18}$&1.19$^{+0.07}_{-0.07}$  & 1.21$^{+0.08}_{-0.07}$ & 1.00 \\
       (Tot)$^{\star}$    & 0.09$^{+0.01}_{-0.01}$ &        --                                                                 & 1.74$^{+0.02}_{-0.02}$     & 93$^{+12}_{-10}$    & 0.49$^{+0.09}_{-0.08}$   &134$^{+25}_{-21}$ &1.09$^{+0.03}_{-0.03}$  &  1.11$^{+0.01}_{-0.01}$ & 0.98 \\    
S5 2116+81      &   --                   &            --                                                              & 1.78$^{+0.05}_{-0.05}$      & $>$93                    & $<$0.64                       & 96$^{+45}_{-44}$&1.14$^{+0.08}_{-0.07}$   & 1.21$^{+0.09}_{-0.08}$ & 0.98 \\
IGR J21247+5058 &           --          & 25.83$^{+17.6}_{-22.2}$  (0.17$^{+0.08}_{-0.09}$)&1.66$^{+0.09}_{-0.10}$ & 96$^{+49}_{-23}$ &    $<$0.06 &23$^{+16}_{-15}$&1.05$^{+0.05}_{-0.05}$    &  1.06$^{+0.05}_{-0.05}$&1.01 \\
                               &                                                &1.69$^{+0.68}_{-0.70}$ ($>$0.81)                           &                                     &                              &                  &                                       &                                         &\\
\hline                 
\hline

\multicolumn{10}{l}{{\footnotesize \emph{Note}: N$_{\rm H}^{\rm FC}$ refers to the column density fully covering the nucleus; N$_{\rm H}^{\rm PC}$ refer to the column density partially covering the nucleus, with cf being the  }} \\
\multicolumn{10}{l}{{\footnotesize  covering  fraction.}}\\
\multicolumn{10}{l}{{\footnotesize    $^{\star}$: sources with Fe line width left as free parameter. MCG+08-11-011: $\sigma$=235$^{+66}_{-65}$ eV; 4U 1344-60: $\sigma$=172$\pm$70 eV; 3C 390.3: $\sigma$=174$^{+110}_{-87}$ eV; }}\\
\multicolumn{10}{l}{{\footnotesize {4C 74.26: $\sigma$=575$^{+110}_{-87}$.}}}\\
\label{fit_res}
\end{tabular}
\end{center}
\end{table*}

\section{Sample extraction and data reduction}
As already done in previous works, we focus on broad line AGN (Seyfert 1-1.5) only, as these are generally less effected 
by spectral features related to absorption; we refer to the sample of objects already analysed
by \citet{Malizia:2014} using non-contemporaneous {\it XMM-Newton} and {\it INTEGRAL}-IBIS/{\it Swift}-BAT data.
For each of these sources we first searched the {\it NuSTAR} data archive for  public observations available up to the end of June 2018 and 
we then considered only those sources with contemporaneous (or quasi-contemporaneous, within a few hours or a day) {\it Swift}/XRT 
observations.
In the case of 4C 74.26 we found four simultaneous {\it XRT/NuSTAR} observations and we analysed each data set individually, as well as their sum in a similar way as done in \citet{Lohfink_2017}.
Sources having simultaneous {\it XMM-Newton} and {\it NuSTAR} observations have been excluded,
due to the fact that being bright objects with very long X-ray observations, their spectra are of high statistical 
quality that cannot be described with a simple phenomenological model, as the one we adopt in the following 
analysis. For this reason, they will be discussed in
detail in a future dedicated paper. 

The final sample is made of 18 AGN which are listed in Table 1 together with the log of all the observations used 
in the analysis. As evident from Table 1, XRT and {\it NuSTAR} observations, even if contemporaneous, have quite different exposures, 
with XRT pointings being generally much shorter (about one third) than the {\it NuSTAR} ones; this might have some consequences when
fitting the broad-band spectra, especially if the source is variable on short timescales (see Section 3).

XRT data (PC mode) reduction for all sources was performed using the standard data pipeline
package (\texttt{XRTPIPELINE} v. 0.13.2) in order to produce screened
event files following the procedure described in \citet{landi10}. Source events were extracted 
within a circular region with a radius of 20 pixels (1 pixel corresponding to 2.36 arcsec) 
centred on the source position, while background events were extracted from a source-free
region close to the X-ray source of interest. 
 For five sources (MCG+08-11-11, IC 4329A, GRS 1734-292, NGC 6814 and IGR J21247+5058) the XRT data were affected by pile-up, i.e. 
the measured rate of the source is high, above about 0.6 counts s$^{-1}$ in the photon-counting mode, and so in these cases spectra were extracted 
using an annular region, thus eliminating the counts in the bright core, where the pile-up will occur.
The spectra were
obtained from the corresponding event files using the \texttt{XSELECT}
v. 2.4c software; we used version v.014 of the response matrices and created 
individual ancillary response files using the task \texttt{xrtmkarf} v.0.6.3.

\emph{NuSTAR} data (from both focal plane detectors, FPMA and FPMB)
were reduced using the \texttt{nustardas\_06Jul17\_V1.8.0} and \texttt{CALDB} version
20180126. For most sources, spectral extraction and the subsequent production of
response and ancillary files was performed using the
\texttt{nuproducts} task with an extraction radius of 75$^{\prime\prime}$; 
to maximise the signal-to-noise ratio, the background spectrum was extracted from a
100$^{\prime\prime}$ radius circular region as close to the source as
possible. 
In the case of GRS 1734-292, due to the presence of stray light 
in the observation\footnote{Note that this source is located in the Galactic Centre region.}, we choose 
an extraction radius of 10 pixels, smaller than what is generally used,
in order to reduce the background contribution, while for the background spectrum a 
radius of 20 pixels has been used, taken on the same detector and including the
same stray light contribution of  the main source.
All parameters in the fits have been allowed to vary unless otherwise stated (see Table 2).
Spectra were generally binned with \texttt{GRPPHA} in order to achieve a minimum of 20 counts per bin so that the $\chi^2$ statistics could be applied.
Spectral fitting was performed in \texttt{XSPEC} v12.9.1m \citep{Arnaud:1996} using $\chi^2$ statistics;
uncertainties are listed at the 90\% confidence level
($\Delta\chi^2$=2.71 for one parameter of interest).

\section{Broad-band spectral analysis}

Spectral fits were carried out in the 0.5-78 keV band,  unless the XRT data did not allow to consider spectral channels below 1 keV, employing the same model for all the sources, i.e. 
the \texttt{pexrav} model in \texttt{XSPEC} (\citealt{Magdziarz_1995}), which consists in an exponentially cut-off
power law reflected from neutral material. Galactic absorption (\texttt{wabs} model in 
\texttt{XSPEC}) has been considered in all fits, {\bf fixing its value to the one from} \citet{Kalberla_2005}.
Additional features are also added whenever they are required by the data  at a significance level greater
than 99.95\%. These extra components are cold and neutral absorption
(\texttt{wabs} model in \texttt{XSPEC}), and or complex, neutral absorption partially covering the central source 
(\texttt{pcfabs} model in \texttt{XSPEC}). Furthermore, in order to account for the excesses around 6 keV a Gaussian K$\alpha$ iron line around 6.4 keV has been added; 
we fixed its width at 10 eV, except in those cases where the sources are known from the literature to have a broad 
K$\alpha$ line or an iron line complex resulting in a blending of two or more lines (\texttt{zgauss} in
\texttt{XSPEC}; see Table 2).
In a few cases, where the data show significant evidence for the presence of a soft excess component, the broad-band 
fit was restricted to the 1-78 keV energy range, since modelling of low energy features is beyond the scope of this paper
and, in any case, the poor statistical quality of the XRT data could not allow an appropriate fit.
However, XRT data are more suitable than the \emph{NuSTAR} ones
to constrain the amount of intrinsic absorption.
Despite the data being simultaneous in most cases, a cross-calibration constant has been 
nonetheless added in the fits, to take into
account differences in the normalisation between XRT and {\it NuSTAR}, but also to adjust for
intrinsic variations in the spectra, either in flux or in spectral shape.

\begin{figure}
\centering
\includegraphics[scale=0.4]{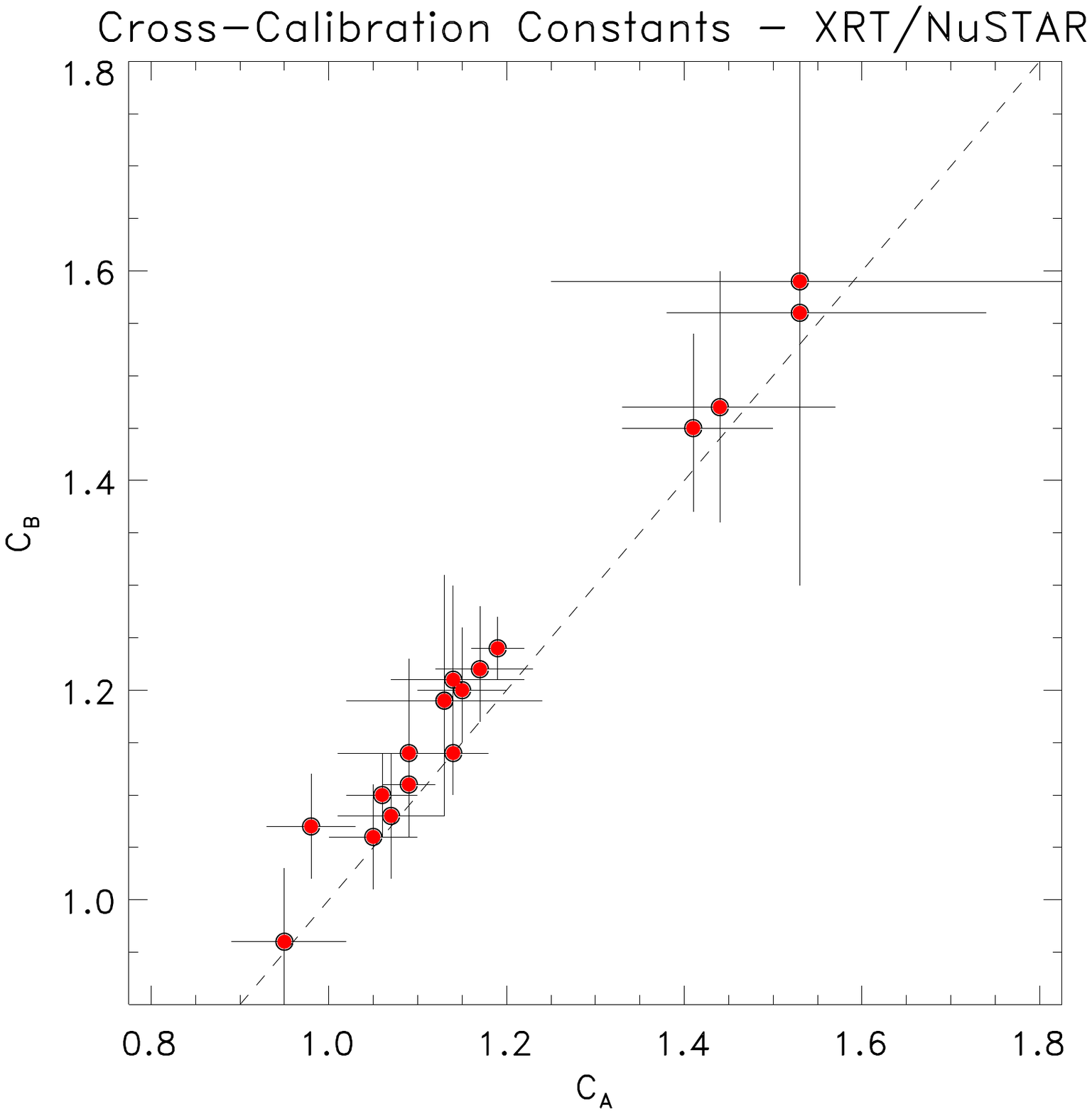}
\caption{{\small Plot of the cross-calibration constants between {\it Swift}/XRT and the
two {\it NuSTAR} detectors (C$_{\rm A}$ and C$_{\rm B}$).}}
\label{const}
\end{figure}

The use of the \texttt{pexrav} model was dictated by the need to compare our results to previous estimates,
but also because this model provides a good approximation of AGN broad-band data. Indeed, the
fit results, reported in Table 2  and the plots reported in Figures \ref{ic_16119}, 
\ref{6814_4c} and \ref{2116_21247} of the Appendix, show that 
this simple phenomenological model describes the data sufficiently well, as demonstrated by the reduced $\chi^2$ reported in the last 
column of the table and by the plot shown in Figure \ref{const}, where the cross calibration constants between 
XRT and the two {\it NuSTAR} detectors are shown. Although most points cluster below 1.2 for both detectors, there are a few objects, i.e. 
QSO B0241+62, Mrk 6, IGR J12415-5750 and IGR J16482-3036, that display higher constant values albeit with larger errors.
In the cases of QSO B0241+62 and Mrk 6 the non perfect simultaneity of the data (see Table 1) and  possibly some variability  
can explain values of the cross-calibration constants around 1.4-1.5. As far as IGR J12415-5750 and IGR J16482-3036 are 
concerned, from the inspection of the light curves, both sources are found to be extremely variable in the 2-10 keV band, therefore 
producing also in these cases constants greater than 1.4 in the broad-band fits. It is also possible that, due to the low statistical 
quality of the soft X-ray data, imperfect modelling at low energies produce a not completely perfect match between 
the XRT/{\it NuSTAR} datasets.
Finally, we note that all our points in Figure \ref{const} are above the 1-1 line, thus reproducing the known effect of a slightly 
different behaviour between {\it NuSTAR} detectors A and B.

\begin{figure}
\centering
\includegraphics[scale=0.4]{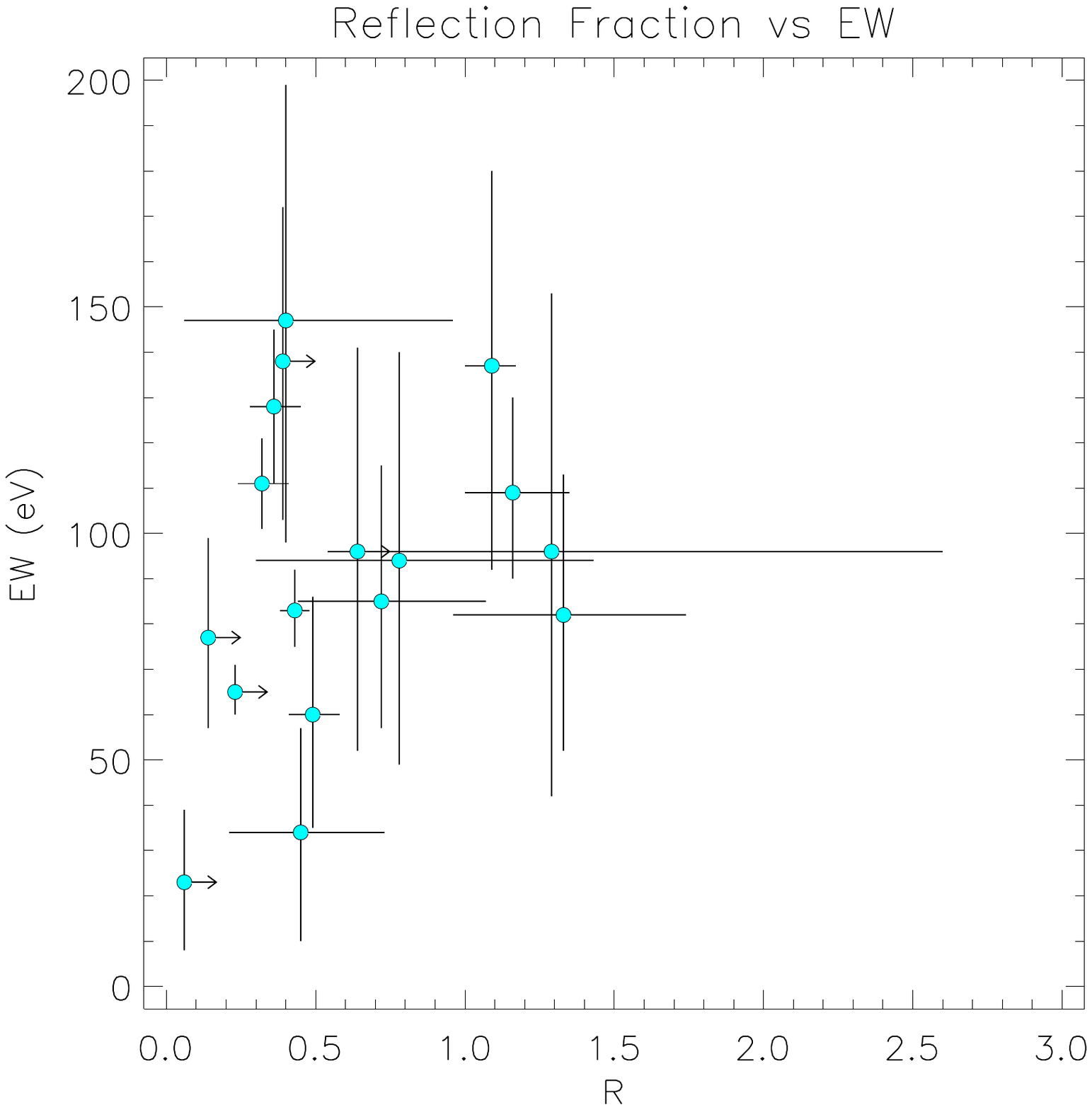}
\caption{{\small Plot of the EW vs reflection fraction (see table for values of the spectral fits).}}
\label{ew_r}
\end{figure}

The mean (arithmetic) value of the photon index estimated for our set of 18 AGN is 1.74, with a dispersion of 0.13, a value which is in 
good agreement with the one  obtained by \citet{Malizia:2014} and by \citet{Ricci_2017} for the same
set of sources. 
The mean (arithmetic) value of the cut-off energy is instead (including lower limits) 111 keV, with a dispersion of 45 keV 
(see next section for a comparison with previous cut-off studies).
In Table 2 we also list the values of the reflection fraction R and of the neutral iron line equivalent width EW: the mean arithmetic value 
of R is found to be (including lower limits) 0.58, with a dispersion of 0.41
while that of EW is 92 eV with a dispersion of 37 eV.
These values are slightly lower than those found in previous works and deserve  a future dedicated analysis  using data with a higher statistical quality, adopting more physical models 
and considering homogeneous samples of AGN (i.e. radio quiet versus radio loud).  If the iron line emission comes from the same medium that produces the reflection, then one would expect to find a 
correlation between these two parameters: in this case the ratio 
between EW over R is typically 100-130 eV \citep{Lubinski_2001}.
This is broadly compatible with the average values quoted above especially considering the
large dispersions found. Despite this, we do not find any correlation between R and EW 
(see Figure \ref{ew_r}) as indicated by the Pearson test which gives
a value of 0.26. Given the uncertainties on both parameters a better correlation test requires higher quality broad-band data as provided by {\it XMM} and {\it NuSTAR}.

\begin{figure}
\centering
\includegraphics[scale=0.4]{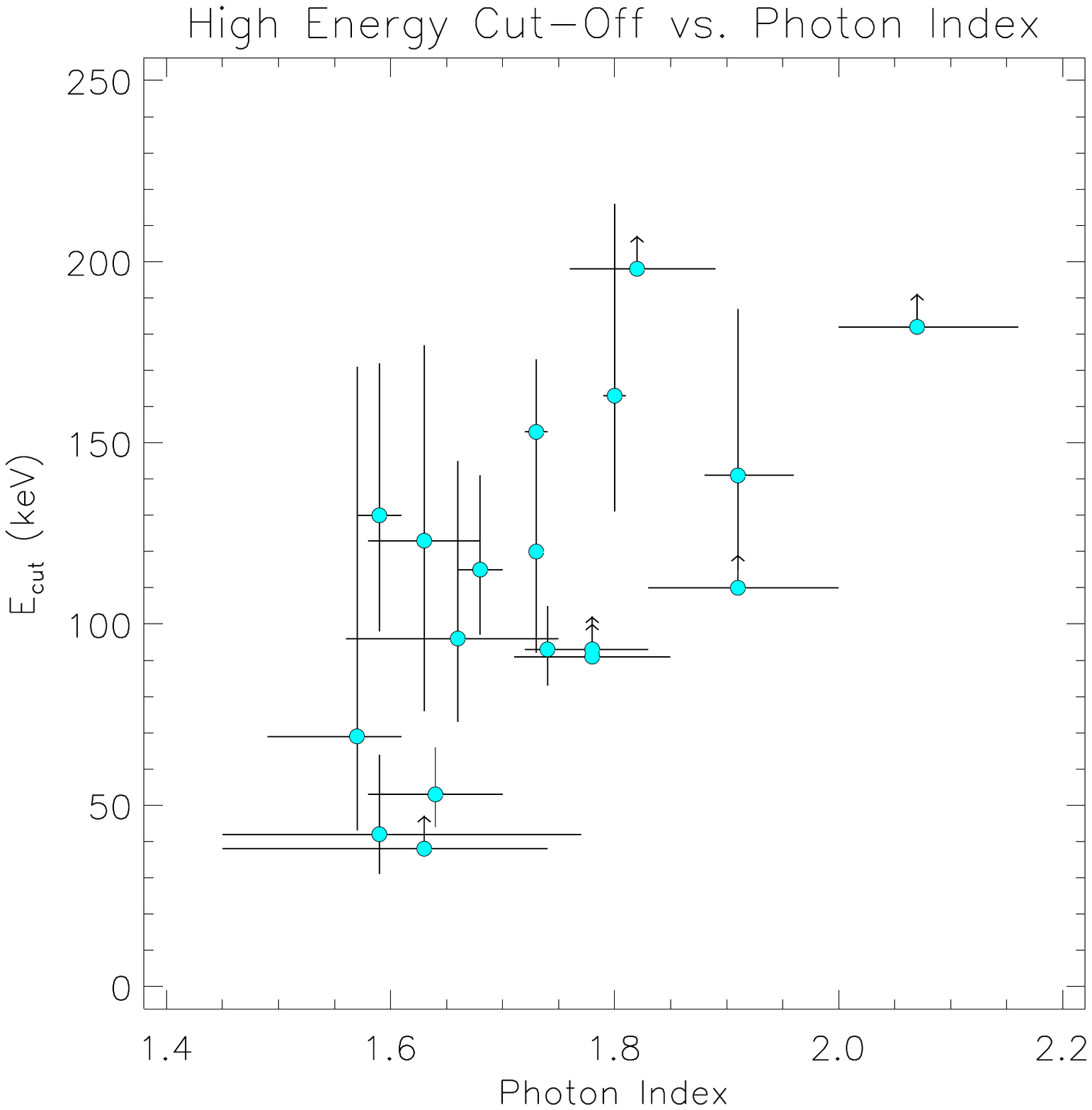}
\caption{{\small High energy cut-off vs. photon index (values form the
spectral fit reported here; see text for details).}}
\label{cutoff}
\end{figure}

Figure \ref{cutoff} shows the photon index {\it vs.} the high energy cut-off obtained from the
\texttt{pexrav} model for the sources analysed in this work. There appears to be a trend of increasing high energy cut-off with 
increasing photon index values, and indeed, if we apply a Pearson statistical test to the data, we find a correlation coefficient of 0.68. 
This trend has already been observed in the past (e.g. \citealt{Perola:2002}), but more recent studies 
(e.g. \citealt{Molina:2009}, \citealt{Tortosa_2018}) have shown 
that no  correlation is seen. 
Given that these two parameters are correlated in the fit
procedure, the weak  correlation observed in the present sample is probably an artifact due to systematic uncertainties on one of the two parameters 
as indeed is found plotting the contours of photon index vs cut-off energy
for three sources chosen over the values range (see Figure A4 of the Appendix.)
This is also expected if the model used 
does not describe the data sufficiently well especially at low energies mainly due to the poor statistical quality of the XRT spectra. This effect can only be
dealt with using high quality, simultaneous X-ray spectra, such as those obtained by {\it XMM-Newton} in combination with 
{\it NuSTAR}, but unfortunately these are available only for a limited number of sources.

\begin{figure}
\centering
\includegraphics[scale=0.4]{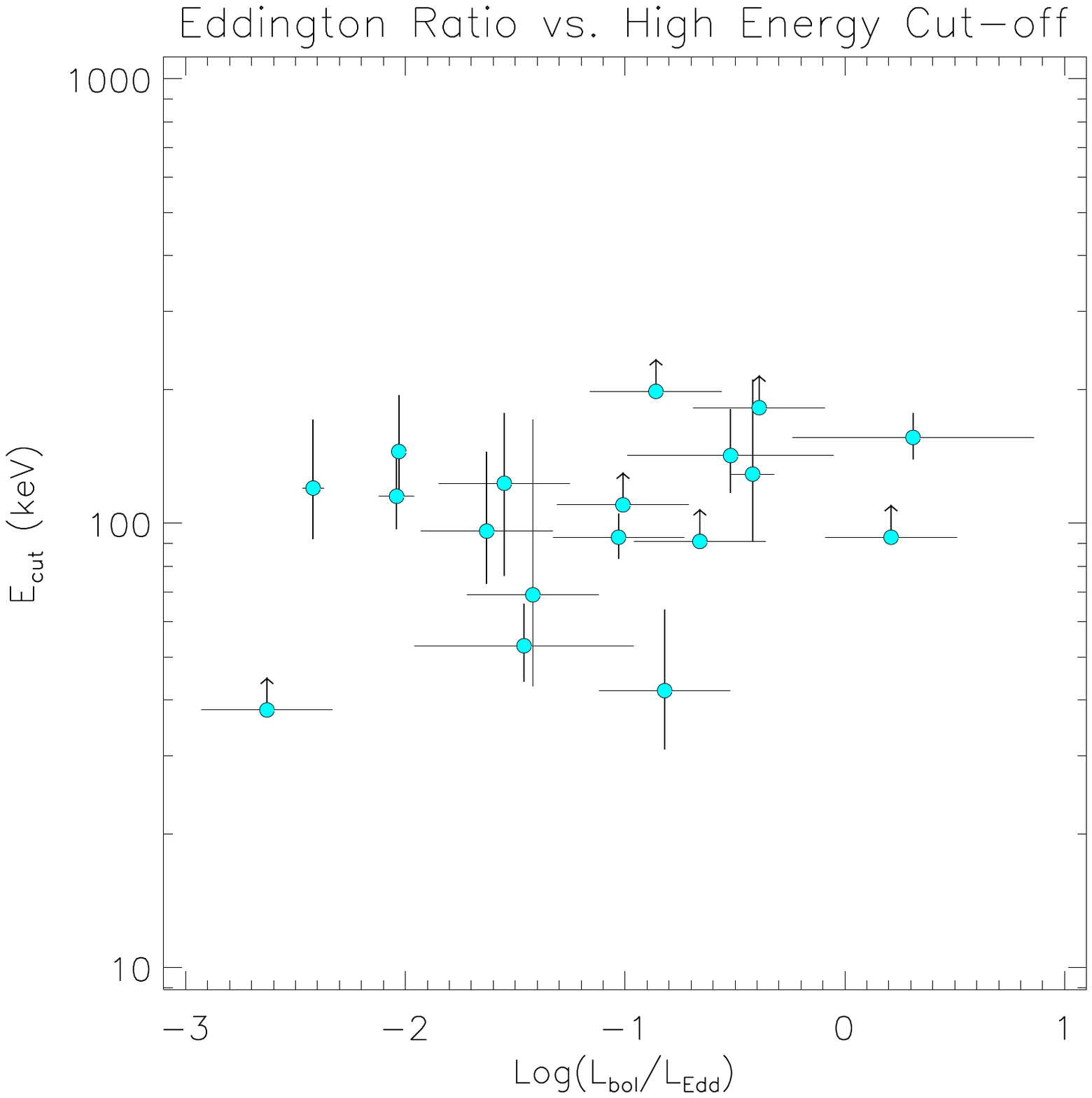}
\caption{{\small High energy cut-off vs. Eddington Ratio for the sources analysed here. }}
\label{edd_ratio}
\end{figure}

Given previous evidence for the decrease of the cut-off energy with increasing
Eddington ratio \citep{Ricci_2018}, we have also searched for a correlation between these two parameters.
Table \ref{accretion} reports for each source in the sample, the 2-10 keV absorption corrected luminosity and the 
bolometric luminosity obtained by applying the correction of Marconi et al. (2003), the most accurate black hole mass found in the 
literature  (see last column of Table \ref{accretion}) and the corresponding Eddington luminosity. The Eddington ratio $\lambda$=L$_{\rm bol}$/L$_{\rm Edd}$ is also listed with its 
relative uncertainty derived from the errors on the black hole mass and 2-10 keV luminosity measurements.
Figure \ref{edd_ratio} shows the high energy cut-off plotted against the Eddington ratios for all the sources in the sample. 
We tested for a possible correlation between the two quantities by applying a Pearson statistical test, but found none, 
being the correlation coefficient 0.34.

Finally, we have also considered the position of our sources in the compactness versus temperature plane as discussed in Fabian et al. (2015, 2017)
and have found that all sources lie in the region allowed by the pair thermostat. To transform the cut-off energy into the coronal temperature kT$_{\rm e}$
we have used a factor of 2. i.e. kT$_{\rm e}$=E$_{\rm cut}$/2, appropriate if $\tau$ (optical depth) is lower than 1 (see \citet{Petrucci:2001}). 
Within this assumption, kT$_{\rm e}$ ranges from 20 keV to 80 keV. However some sources, like IGR J16119-6036, 
2E 1853.7+1534 and GRS 1734-292 have temperatures which are too low to be compatible with a pure thermal plasma, 
introducing the need for a hybrid plasma where a non-thermal component, which tends to lower the limiting temperature, becomes significant (\citealt{Fabian_2017}). 
Particularly in these cases, simultaneous {\it XMM}/{\it NuSTAR} observations would be helpful in 
confirming these low kT$_{e}$ values, which would test the current models for a hybrid plasma.

\begin{table*}
\small
\begin{center}
\vspace{0.2cm}
\begin{tabular}{lcccccl}
\multicolumn{7}{c}{{\bf Table 3 - \texttt Accretion Parameters}} \\
\hline
{\bf Source} 	&{\bf Log(L$_{\rm \bf 2-10 keV}$)}& {\bf Log(L$_{\rm \bf Bol}^{\rm \bf corr}$)}&{\bf Log(L$_{\rm \bf Edd}$)}&{\bf Log($\bf \lambda$)}&{\bf Log(M$_{\rm \bf BH}$)} &  {\bf Ref.}\\
\hline
QSO 0241+62     & 43.85$\pm$0.04    & 45.33	&       46.19	&	-0.860$\pm$0.30        &         8.09$\pm$0.30 & \citet{Koss_2017}    \\
MCG+08-11-011   & 43.62$\pm$0.01    & 45.03	&	45.55	&	-0.520$\pm$0.47        &	 7.45$\pm$0.47 & \citet{Fausnaugh_2017}     \\
Mrk 6           & 42.65$\pm$0.03    & 43.81	&	46.23	&	-2.420$\pm$0.05        &	 8.13$\pm$0.04 & \citet{Grier_2012}     \\
FRL 1146        & 43.45$\pm$0.02    & 44.81	&	45.20	&	-0.390$\pm$0.30        &	 7.10$\pm$0.3  & ($\star$)     \\
IGR J12415-5750 & 43.44$\pm$0.02    & 44.80	&	46.35	&	-1.550$\pm$0.30        &	 8.25$\pm$0.3  & \citet{Koss_2017}      \\
4U 1344-60      & 43.13$\pm$0.03    & 44.41	&   $<$46.44& $>$-2.030$\pm$0.03       &	 $<$8.34       &  ($\star$ $\star$)   \\
IC 4329A	    & 43.73$\pm$0.02    & 45.18 &	44.87	&	 0.310$\pm$0.55        &	 6.77$\pm$0.55 & \citet{Kaspi_2000}    \\
IGR J16119-6036 & 42.75$\pm$0.02    & 43.94	&	45.36	&	-1.420$\pm$0.30        &	 7.26$\pm$0.30 & \citet{Koss_2017}  \\
IGR J16482-3036 & 42.50$\pm$0.05    & 43.62	&	46.25	&	-2.630$\pm$0.30        &	 8.15$\pm$0.30 & \citet{Masetti_2006}    \\
GRS 1734-292	& 43.70$\pm$0.02    & 45.14	&   46.60	&	-1.460$\pm$0.50        &	 8.50$\pm$0.50 & \citet{Tortosa:2017}     \\
2E 1739.1-1210  & 43.50$\pm$0.02    & 44.88	&   45.89	&	-1.010$\pm$0.30        &	 7.79$\pm$0.3  & \citet{Koss_2017}           \\
IGR J18027-1455 & 43.75$\pm$0.02    & 45.20	&   45.86	&	-0.660$\pm$0.30        &	 7.76$\pm$0.3  & \citet{Koss_2017}         \\
3C 390.3        & 44.41$\pm$0.02    & 46.07	&	46.49	&	-0.420$\pm$0.10        &	 8.39$\pm$0.10 & \citet{Feng_2014}         \\
2E 1853.7+1534  & 44.04$\pm$0.01    & 45.57	&	46.39	&	-0.820$\pm$0.30        &	 8.29$\pm$0.3  & \citet{Koss_2017}          \\
NGC 6814	    & 42.25$\pm$0.01    & 43.32	&	45.36	&	-2.040$\pm$0.08        &	 7.26$\pm$0.08 & \citet{Feng_2014}         \\
4C 74.26	    & 44.86$\pm$0.01    & 46.67	&	47.70	&	-1.030$\pm$0.30        &	 9.60$\pm$0.3  & \citet{Koss_2017}       \\
S5 2116+81      & 44.31$\pm$0.02    & 45.94	&   45.73	&	 0.210$\pm$0.30        &	 7.63$\pm$0.3  & \citet{Koss_2017}          \\
IGR J21247+5058 & 43.81$\pm$0.02    & 45.27	&   46.90	&	-1.630$\pm$0.30        &	 8.80$\pm$0.3  & \citet{Koss_2017}          \\

\hline
\label{lum_mass}
\end{tabular}
\end{center}
Notes: ($\star$): estimated from \citet{Rodriguez_Ardila_2000} data, using the formula of \citet{Khorunzhev_2012};
($\star$ $\star$) estimated from \citet{Masetti_2006}  data, using the formula of \citet{Greene_2005};
if not reported in the reference paper, errors on the black hole masses have been assumed to be 0.3 dex for broad  broad H$_\beta$ measurements.

\label{accretion}
\end{table*}

\section{The high energy cut-off: comparison with previous works}

In table \ref{ec_compare},  we report the cut-off energies obtained for the same set of sources by \citet{Malizia:2014} 
and \citet{Ricci_2017} where similar modelling of the broad-band spectra were employed, so that a straightforward comparison is possible. 
For completeness, we also report the cut-off energy obtained from {\it NuSTAR} data available in the literature (see table for references).
Interestingly, out of 18 objects analysed in this work, only 7 have a published {\it NuSTAR} measurement on the cut-off energy; 
for the remaining sources, we have been able to measure for the first time a high energy cut-off in five and set upper limits in six objects.

Figure \ref{compare}, left panel, shows the values of the high energy cut-off found in the present work versus the ones reported 
in \citet{Malizia:2014}. Within 2$\sigma$ errors, the values obtained from the spectral fitting of simultaneous datasets are
comparable with those obtained from non-simultaneous XMM/IBIS/BAT spectra, therefore
confirming these previous results. This agreement is evident in the right panel of Figure \ref{compare}, where the histogram of the ratios 
between the high energy cut-off values measured in this work and those obtained by \citet{Malizia:2014} is plotted: a clustering of these 
ratios is evident around 1, indicating a good match between 
the two studies. Finally we note that in some cases {\it NuSTAR} data allow to put more stringent constraints on the cut-off energy, while
in other cases, such as those of Fairall 1146 and IGR J16482-3036, only loose constraints could be placed on the parameter, 
referring to \citet{Malizia:2014} for a  more  accurate estimate of the high energy cut-off.

Figure \ref{histo_compare}, left panel, shows instead the values of the cut-off energy for the sources in our sample compared
with the results reported in \citet{Ricci_2017} and with previous studies performed using {\it NuSTAR} data. 
We note that in this case there is a lesser agreement between the high energy cut-off values 
measured in the two studies and a tendency for the Ricci et al. measurements to be systematically higher than what we find in the present analysis.
Again this is highlighted in the right panel of Figure \ref{histo_compare}, where we plot the histogram of the ratio between the high energy
cut-off values measured in this work and those obtained by Ricci et al.: in this case a similar clustering is evident, but around 0.5
indicating a systematic mismatch between the two measurements. 
It is worth noting that {\it NuSTAR} data employed in this work allow to put better constraints than those reported in \citet{Ricci_2017}
on the high energy cut-off in almost all cases, except for QSO B0241+62.

\begin{figure}
\centering
\includegraphics[scale=0.33, angle =-90]{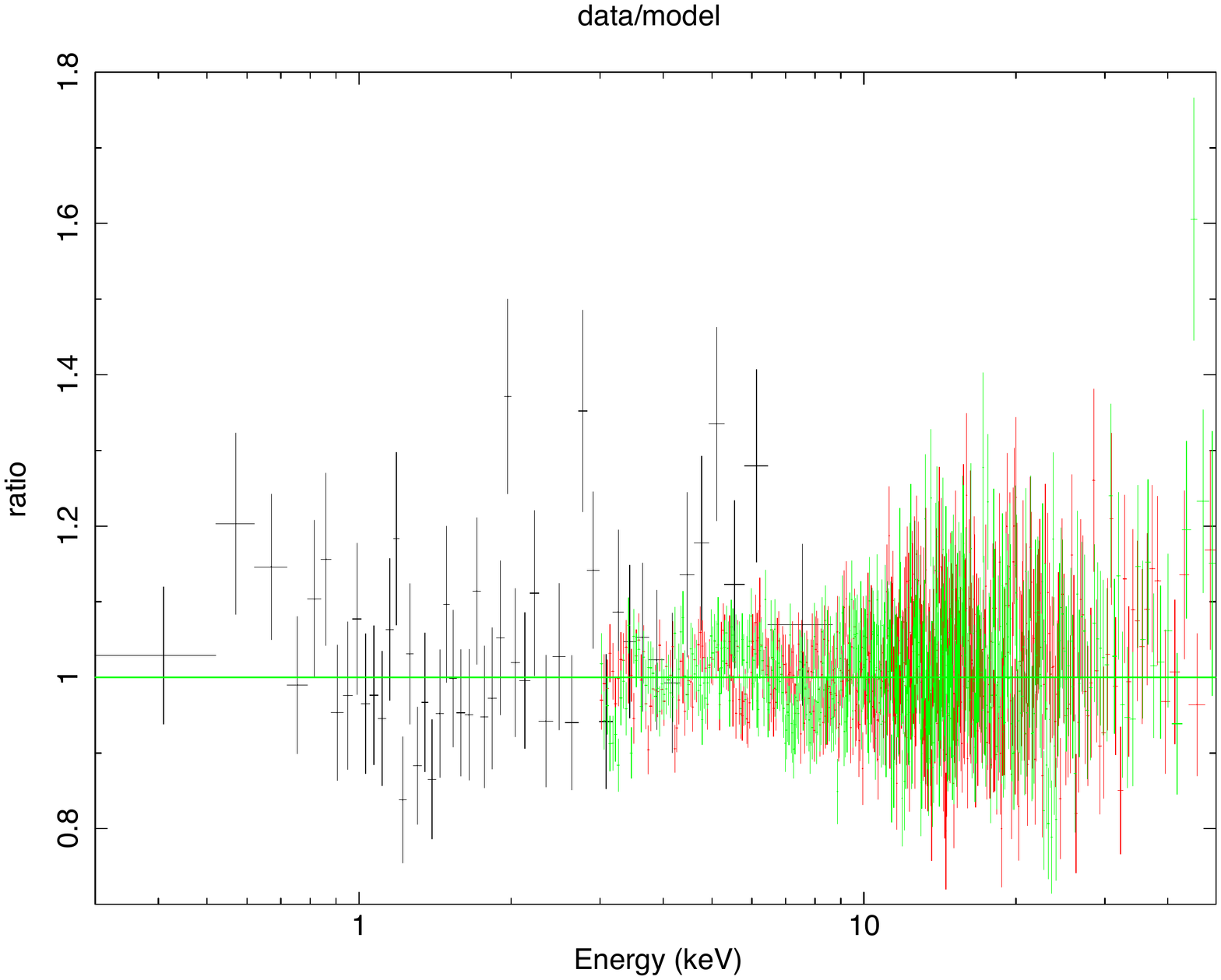}
\caption{{\small Data-to-model ratio for the summed observations of 4C 74.26 employing the \texttt{relxilllp} model, following \citet{Lohfink_2017}.}}
\label{4c_ratio}
\end{figure}

Finally, a comparison with previous {\it NuSTAR} studies of 7 sources in the sample indicates
a perfect match with our results, 
except for the case of 4C 74.26, where the cut-off values reported in the literature 
\citep{Lohfink_2017} are higher than what we find in this work. Interestingly, Malizia et al.
(2014) 
report similar values as those obtained by \citet{Lohfink_2017}. This source is known to host
a photoionised outflow \citep{Di_Gesu_2016} and both \citet{Lohfink_2017} and  
\citet{Malizia:2014} took this component into account in their modelling of the source, while
we were unable with the XRT data to reproduce it. Furthermore 
\citet{Lohfink_2017} employ a different  broad-band model (\texttt{relxilllp}) to fit the 
source spectrum. We have therefore re-analized our data to take into account both ionized 
reflection and warm absorber in the fit, but the high energy cut-off resulted always in a 
lower value, either in the individual observations or in their sum, in agreement with our
initial  
analysis. We have also re-extracted the data using the same procedure details adopted by
\citet{Lohfink_2017}, mplementing the exact same model, i.e. \texttt{relxilllp} but could not
reproduce their results, in fact the high energy cut-off remains lower than
a factor of two, but  the fit is nonetheless good ($\chi^2$=1814 for 1800 d.o.f.; see Figure \ref{4c_ratio}). 
At this point, the reasons behind this discrepancy are not clear.

Overall, we can conclude that our analysis confirms previous {\it NuSTAR} studies of most of our sources and furthermore indicates that a simple phenomenological
model like \texttt{pexrav}, provides a good fit of  broad-band spectra when data are not of high statistical quality. Moreover, the good match obtained comparing the present results
 with those of
\citet{Malizia:2014}, strongly suggests that non simultaneity is not a major issue; this is a useful information that allows the use of good quality 
average spectra such as those accumulated by {\it INTEGRAL}/IBIS over a large number of AGN.

\begin{table*}
\small
\begin{center}
\vspace{0.2cm}
\begin{tabular}{lccc}
\multicolumn{4}{c}{{\bf Table 4 - \texttt Previous Results}} \\
\hline
{\bf Source}      & {\bf Malizia et al. 2014} & {\bf Ricci et al. 2017}  & {\bf NuSTAR} (Ref)     \\
\hline
QSO B0241+62      & $>$138                    & 177$^{+66}_{-42}$        &    -                   \\
MCG+08-11-011     & 171$^{+44}_{-30}$         & 252$^{+131}_{-60}$       & 175$^{+110}_{-50}$ (1) \\
Mrk 6             & 131$^{+132}_{-48}$        & 122$^{+42}_{-15}$        &    -                   \\
FRL 1146          & 84$^{+79}_{-30}$          & $>$72                    &    -                   \\
IGR J12415-5750   & 175$^{+296}_{-74}$        & $>$229                   &    -                   \\
4U 1344-60        & $>$110                    & 45$^{+7}_{-4}$           &    -                   \\
IC 4329A          & 152$^{+51}_{-32}$         & 236$^{+42}_{-26}$        & 185$^{+15}_{-15}$ (1)  \\
IGR J16119-6036   & $>$100                    & 127$^{+333}_{-64}$       &    -                   \\
IGR J16482-3036   & 163 $^{+220}_{-69}$       & $>$90                    &    -                   \\
GRS 1734-292      & 58$^{+24}_{-7}$           & 84$^{+38}_{-26}$         & 53$^{+10}_{-10}$ (1)    \\
2E 1739.1-1210    & -                         & $>$230                   &    -                    \\
IGR J18027-1455   & $>$86                     & $>$74                    &    -                    \\
3C 390.3          & 97$^{+20}_{-11}$          & 166$^{+64}_{-37}$        & 120$^{+20}_{-20}$ (1)   \\
2E 1853.7+1534    & 89$^{+50}_{-26}$          & 152$^{+413}_{-78}$       &    -                    \\
NGC 6814          & 190$^{+185}_{-66}$        & $>$210                   & 135$^{+70}_{-35}$ (1)   \\ 
4C 74.26 (tot)    & 189$^{+171}_{-66}$        & $>$119                   & 183$^{+51}_{-35}$ (2)   \\
S5 2116+81        & $>$180                    & $>$175                   &      -                  \\
IGR J21247+5058   & 79$^{+23}_{-15}$          & 206$^{+111}_{-27}$       & 100$^{+90}_{-30}$  (3)  \\
\hline
\multicolumn{4}{c}{{\scriptsize References: (1): \citet{Tortosa_2018}; (2) \citet{Lohfink_2017}; (3) \citet{Buisson_2018}.}}
\end{tabular}
\label{ec_compare}
\end{center}
\end{table*}

\begin{small}
\begin{figure*}
\centering
\includegraphics[scale=0.4]{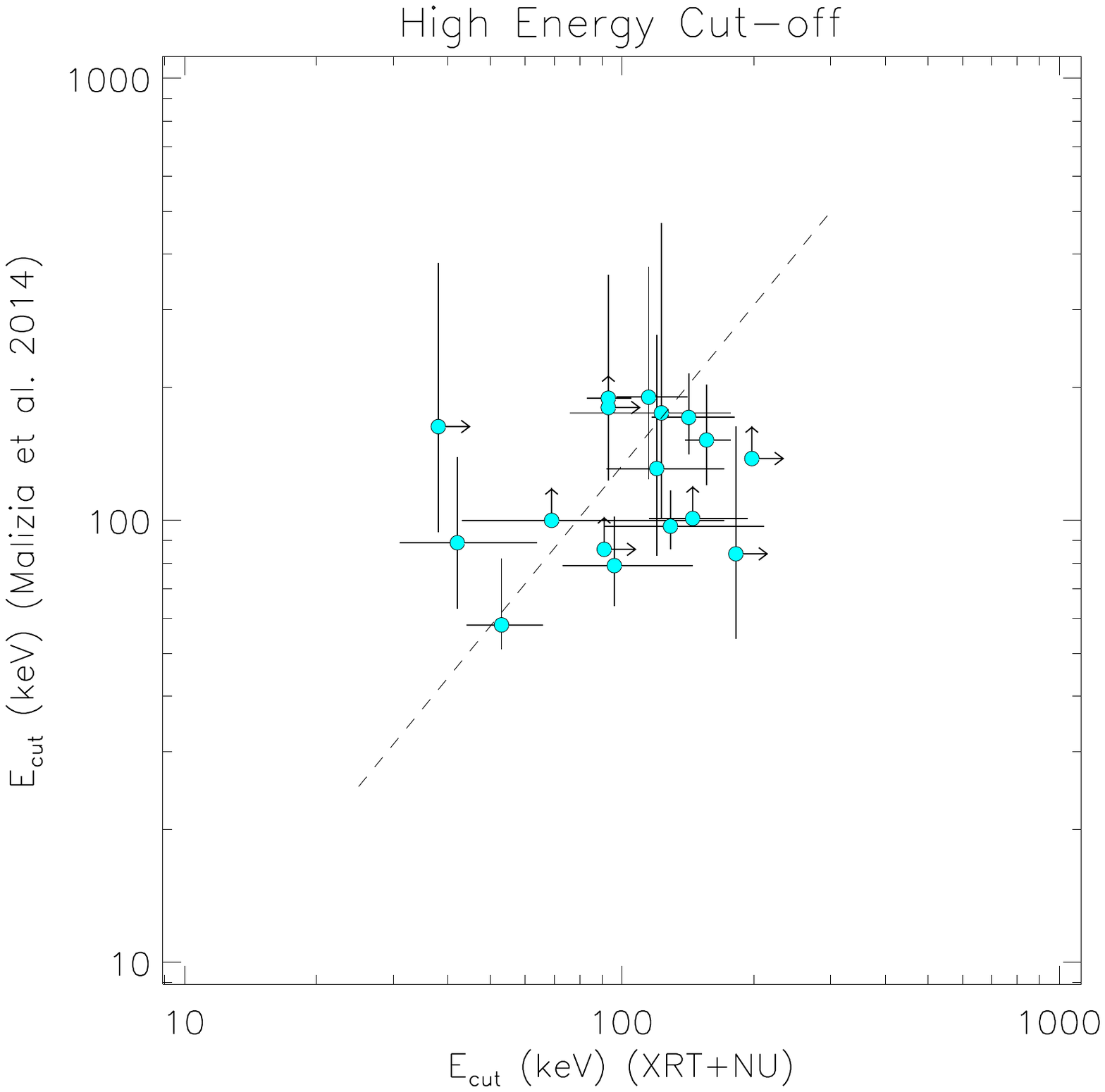}
\hspace{0.5cm}
\includegraphics[scale=0.4]{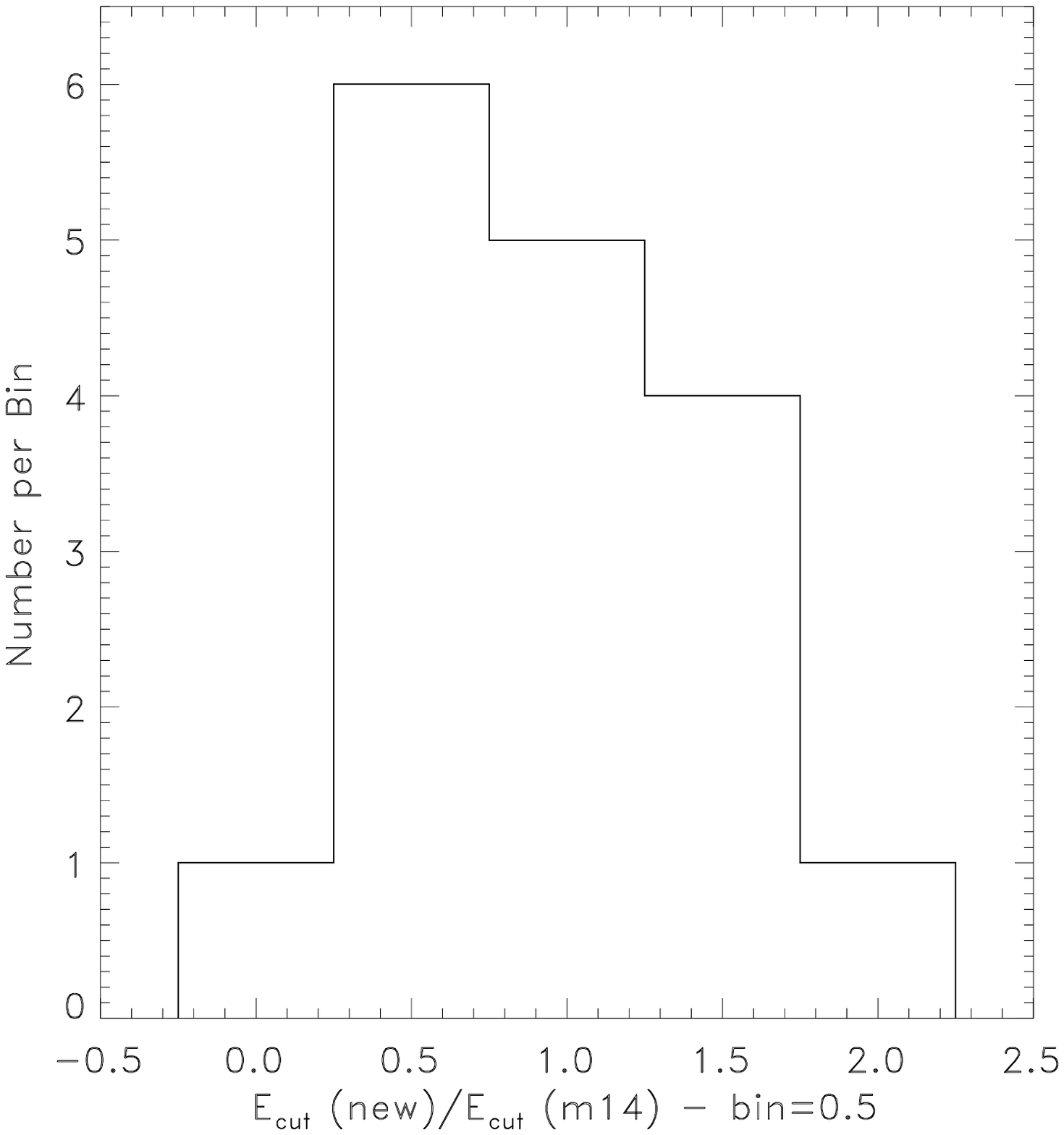}
\caption{{\small {\it Left Panel:} plot of the values of the high energy cut-off from the present analysis plotted against the results from
\citet{Malizia:2014}; the 1:1 line is shown for reference. {\it Right Panel:} histogram of the ratio between the high energy cut-off values 
measured in this work and those obtained by \citet{Malizia:2014}. }}
\label{compare}
\end{figure*}
\end{small}

\begin{small}
\begin{figure*}
\centering
\includegraphics[scale=0.4]{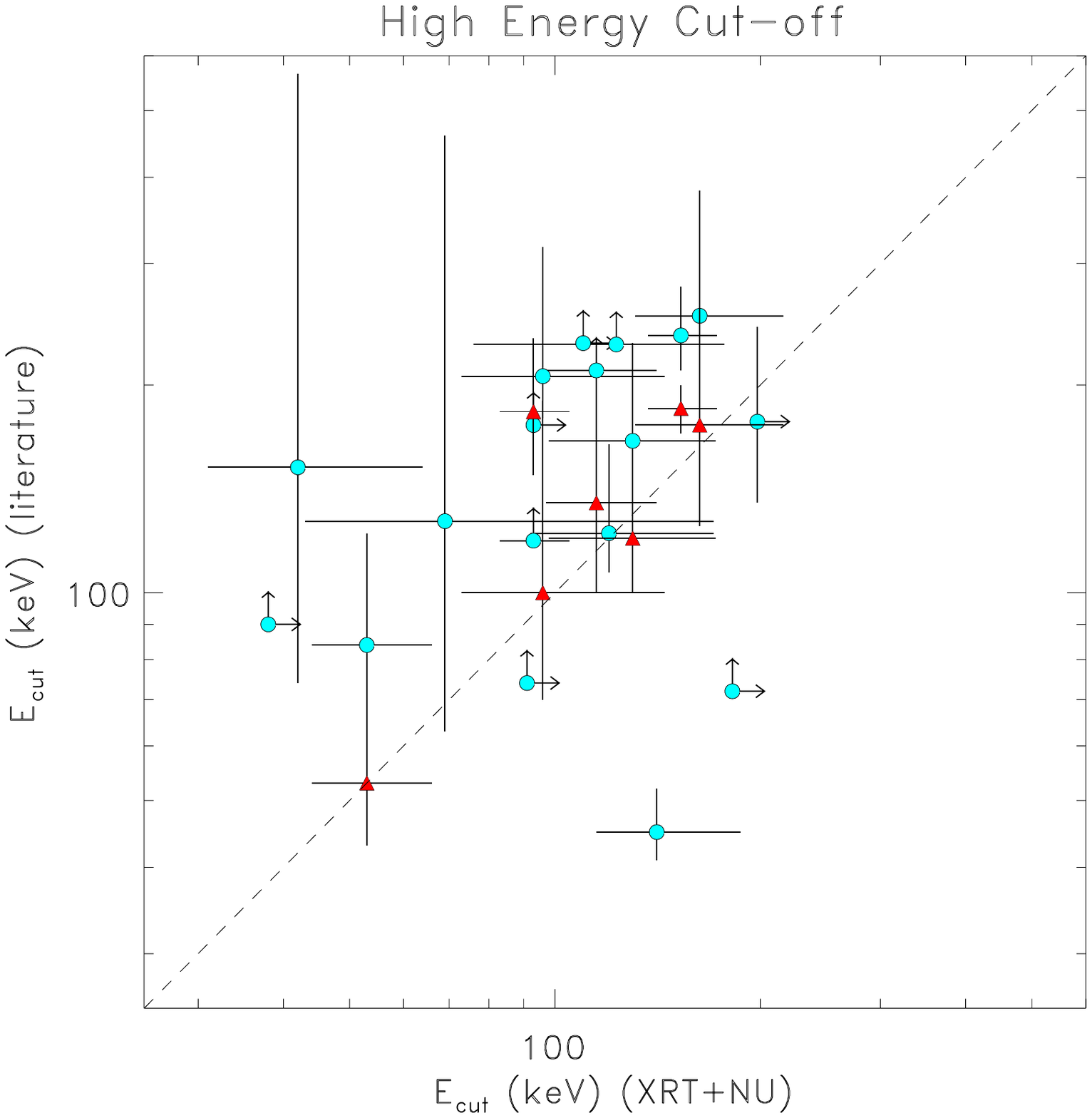}
\hspace{0.5cm}
\includegraphics[scale=0.4]{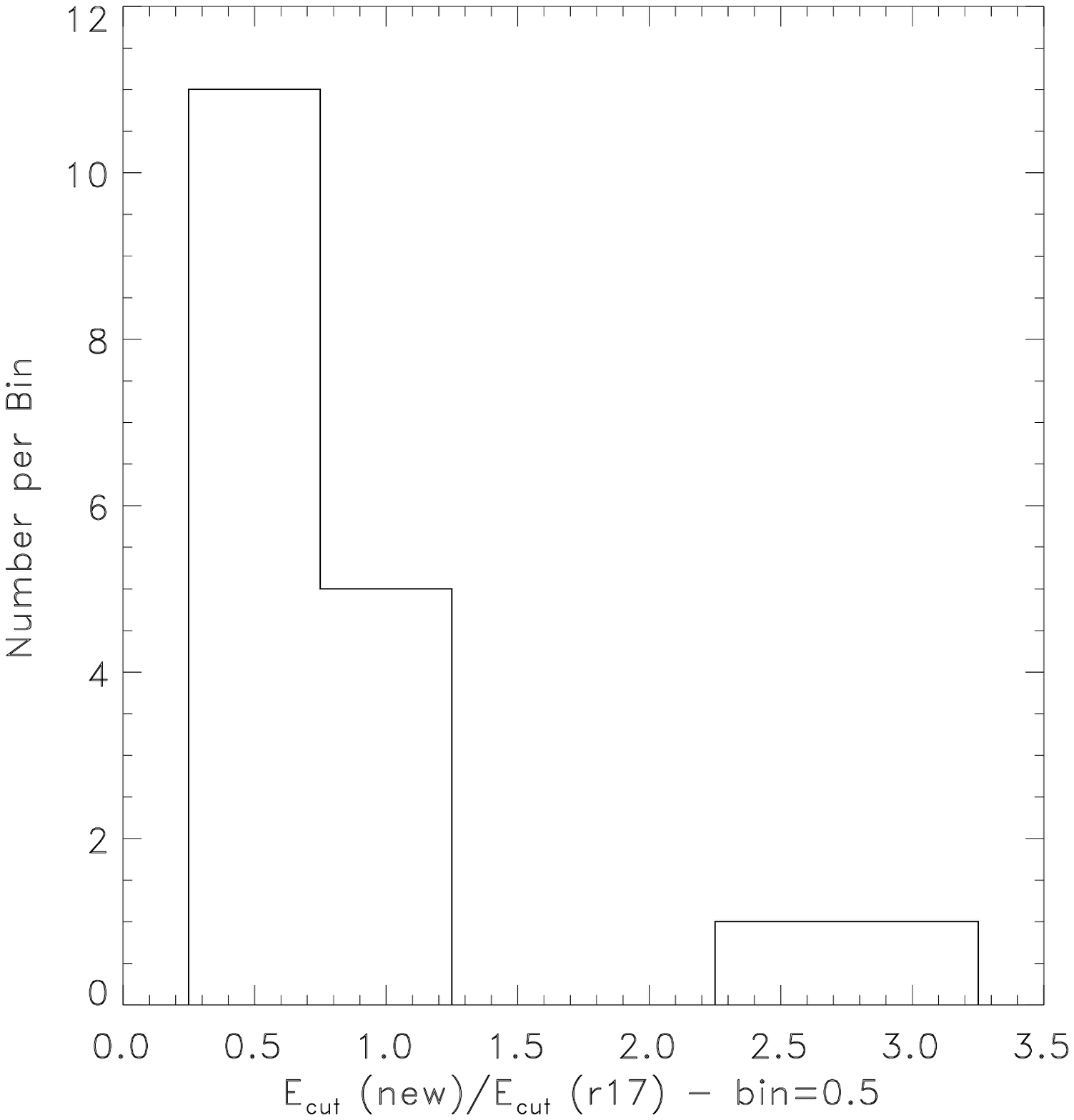}
\caption{{\small {\it Left Panel:} plot of the values of the high energy cut-off from the present analysis plotted against the 
results from \citet{Ricci_2017} (cyan circles) and from several studies employing solely {\it NuSTAR} data (red triangles); the 1:1 line is shown for reference. {\it Right Panel:} 
histogram of the ratio between the high energy cut-off values 
measured in this work and those obtained by \citet{Ricci_2017}.}}
\label{histo_compare}
\end{figure*}
\end{small}

\section{Summary and conclusions}
In this work we have performed the broad-band (0.5-78 keV) spectral analysis of 18 broad line AGN, i.e. all those belonging to the 
{\it INTEGRAL} complete sample of AGN (Malizia et al. 2014) for which contemporaneous {\it Swift}-XRT and {\it NuSTAR} observations were 
available from the archives. Out of the 18 sources analysed, we found a good constraint on the high energy cut-off for 13 objects, 7 of
which have already published {\it NuSTAR} measurements that we confirm in the present study; for the remaining 5 sources, we were able to
place only lower limits on the value of the cut-off energy.
We met the goal of our work by confirming that the distribution of the high energy cut-off in broad line AGN peaks at around 100 keV, 
by analysing simultaneous XRT and {\it NuSTAR} data and we also confirm what was found in our previous work \citep{Malizia:2014}, 
which made use of non-contemporaneous XMM-{\it Newton}, {\it INTEGRAL}-IBIS and {\it Swift}-BAT spectra; 
we found an equally good agreement in the distribution of photon indices, for which we find a value of 1.74.
This is an important result indicating that variability, if any is present, does not influence the spectral fits when dealing with high 
energy data obtained over long timescales as those provided by {\it INTEGRAL}-IBIS and {\it Swift}-BAT.  
Moreover, it is worth noting that both {\it INTEGRAL} and {\it Swift}, having accumulated long exposures on almost the whole sky, are competitive with 
the much more sensitive instruments on board {\it NuSTAR} whose observations typically last a few tens of ksec. 
In order to be able to compare these results with the old ones, we employed the same phenomenological model, \texttt{pexrav}, for all the 
sources as in 2014 and a second important result came by: using relatively simple phenomenological models over more complex and more 
physical ones, provides accurate measurements of the spectral parameters also with datasets of good statistical quality such as  those 
provided by {\it NuSTAR}.
A further confirmation of this is the fact that our cut-off estimates are in perfect agreement with those already published using 
the same {\it NuSTAR} datasets (see Table 4).

Following what was previously done by a number of other studies (e.g. \citealt{Ricci_2018}; \citealt{Tortosa_2018}), we have searched
for correlations in our parameter space. In particular, we tested the much debated correlation between the photon index and the
high energy cut-off \citep{Perola:2002}: we do find a weak correlation between the two quantities, but we ascribe it to a non perfect 
modelisation of the low energy data due to the poor statistical quality of the XRT spectra. We also tested for a possible correlation
between the high energy cut-off and the Eddington ratio, but contrary to what found by \citet{Ricci_2018} and in agreement
with what reported by \citet{Tortosa_2018}, we do not observe any. 

In conclusion, the present study strongly supports the evidence that the high energy cut-off is low and located around 100 keV, using both 
simultaneous and non-simultaneous data, and employing either physical or phenomenological models.
Clearly, to test theoretical predictions on the physical characteristics of the X-ray emitting plasma in the corona and its relation with the properties of the accreting SMBH,
 high  statistical quality broad band data such as those provided by {\it XMM-Newton} in combination with  {\it NuSTAR} and/or {\it INTEGRAL}-IBIS/{\it Swift}-BAT are essential.

\section*{Acknowledgements}
We thank Dr. Matteo Perri (ASDC) for helping in the reduction
of NuSTAR data of GRS 1734-292. M.M. acknowledges financial
support from ASI under contract INTEGRAL ASI/INAF 2013-025-RL and
NuSTAR NARO16 1.05.04.95.




\bibliographystyle{mnras}
\bibliography{mol_biblio} 


\newpage
\appendix
\section{}  in Figures \ref{ic_16119}, \ref{6814_4c} and \ref{2116_21247} are reported the
best fit models together with the model-to-data ratio for the 
sources analysed in the sample. As described in text, we employed the \texttt{pexrav}
model in all sources plus any other component (such as the iron Gaussian line or intrinsic absorption) as required by the data. In Figure \ref{cont} we report, as an example the 
confidence contour plot of the high energy cut-off vs. the photon index for three of the 
sources in the sample, namely MCG+08-11-011, 3C 390.3 and NGC 6814.

\begin{small}
\begin{figure*}
\centering
\includegraphics[scale=0.25,angle=-90]{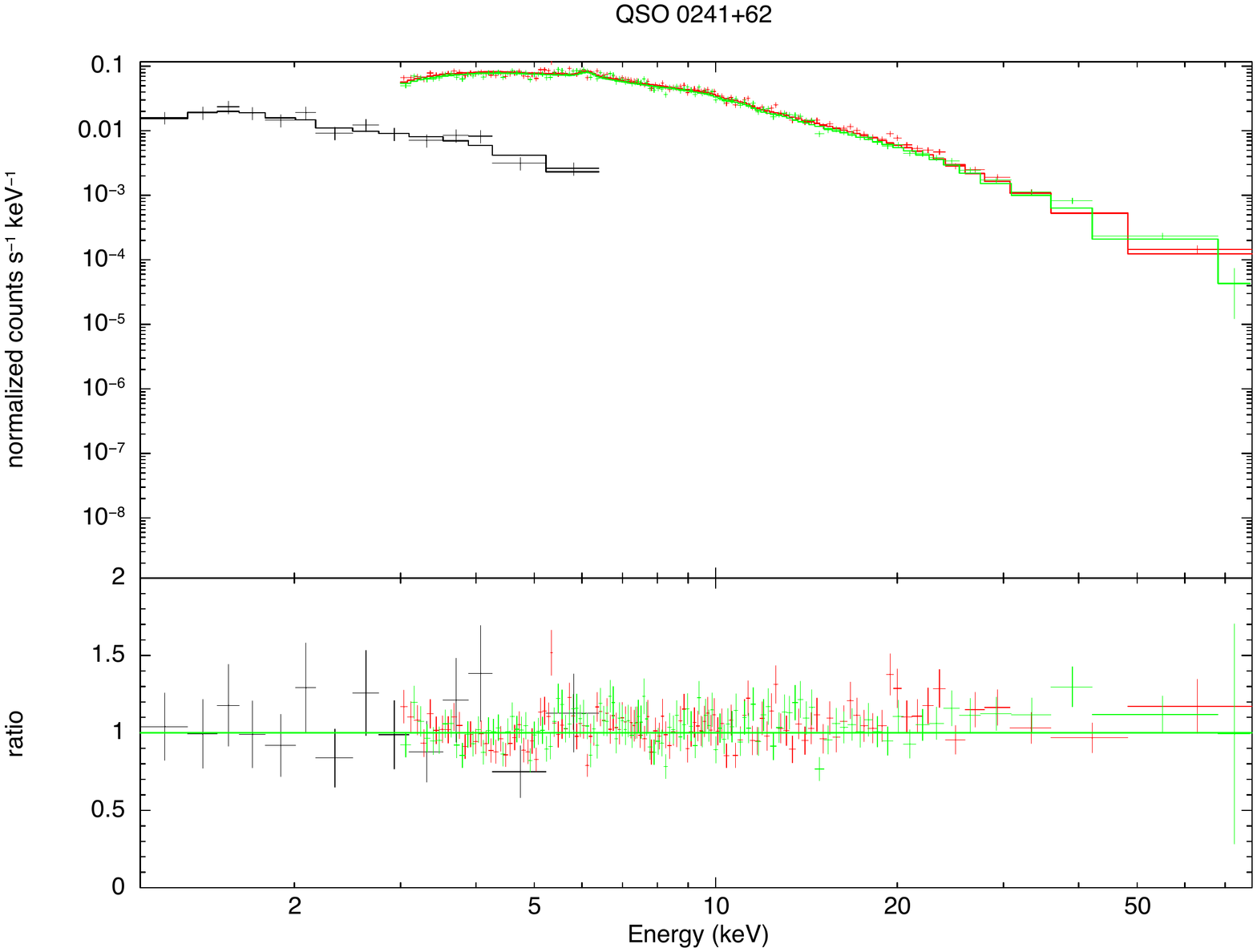}
\hspace{0.4cm}
\includegraphics[scale=0.25,angle=-90]{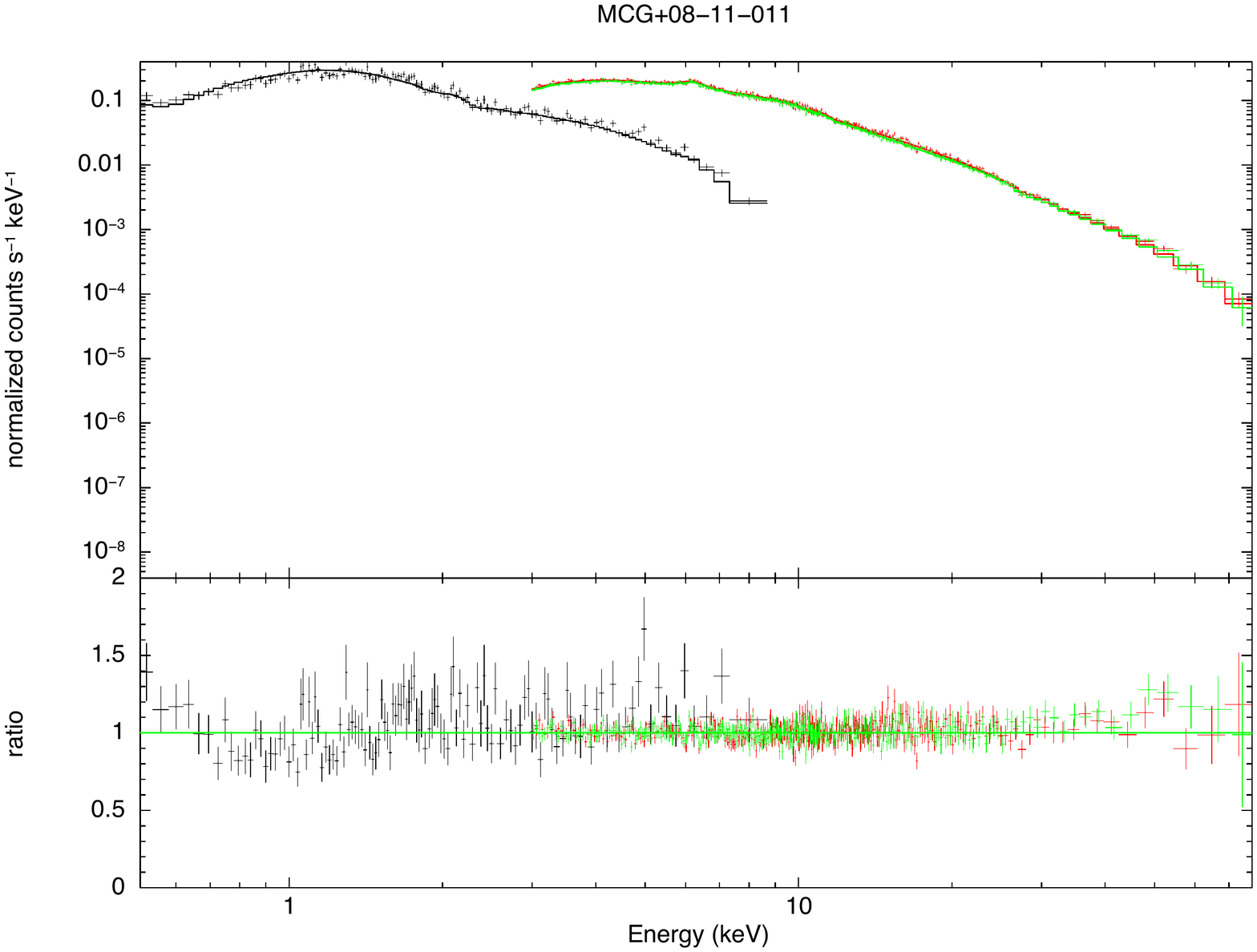}\\
\vspace{0.5cm}
\includegraphics[scale=0.25,angle=-90]{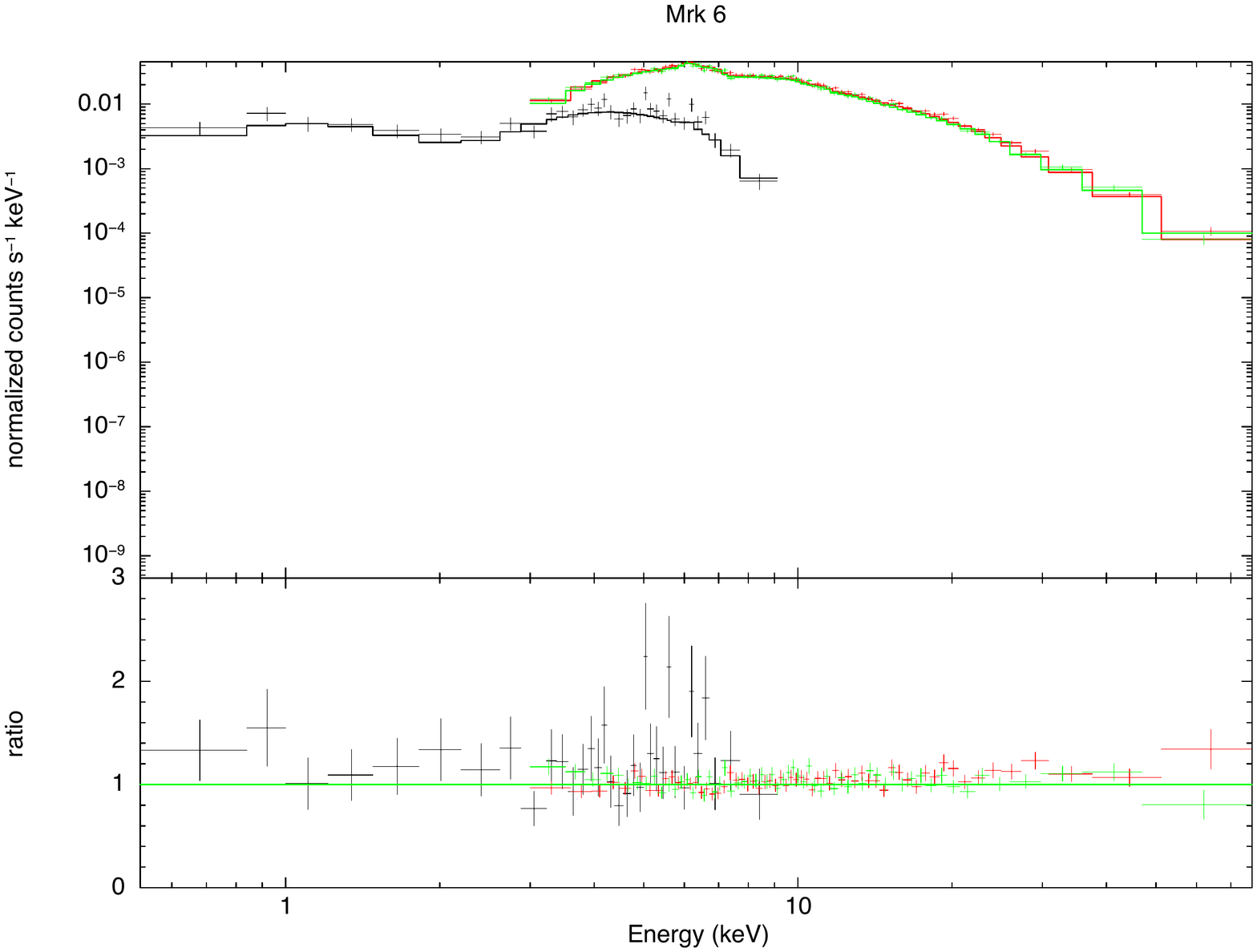}
\hspace{0.4cm}
\includegraphics[scale=0.25,angle=-90]{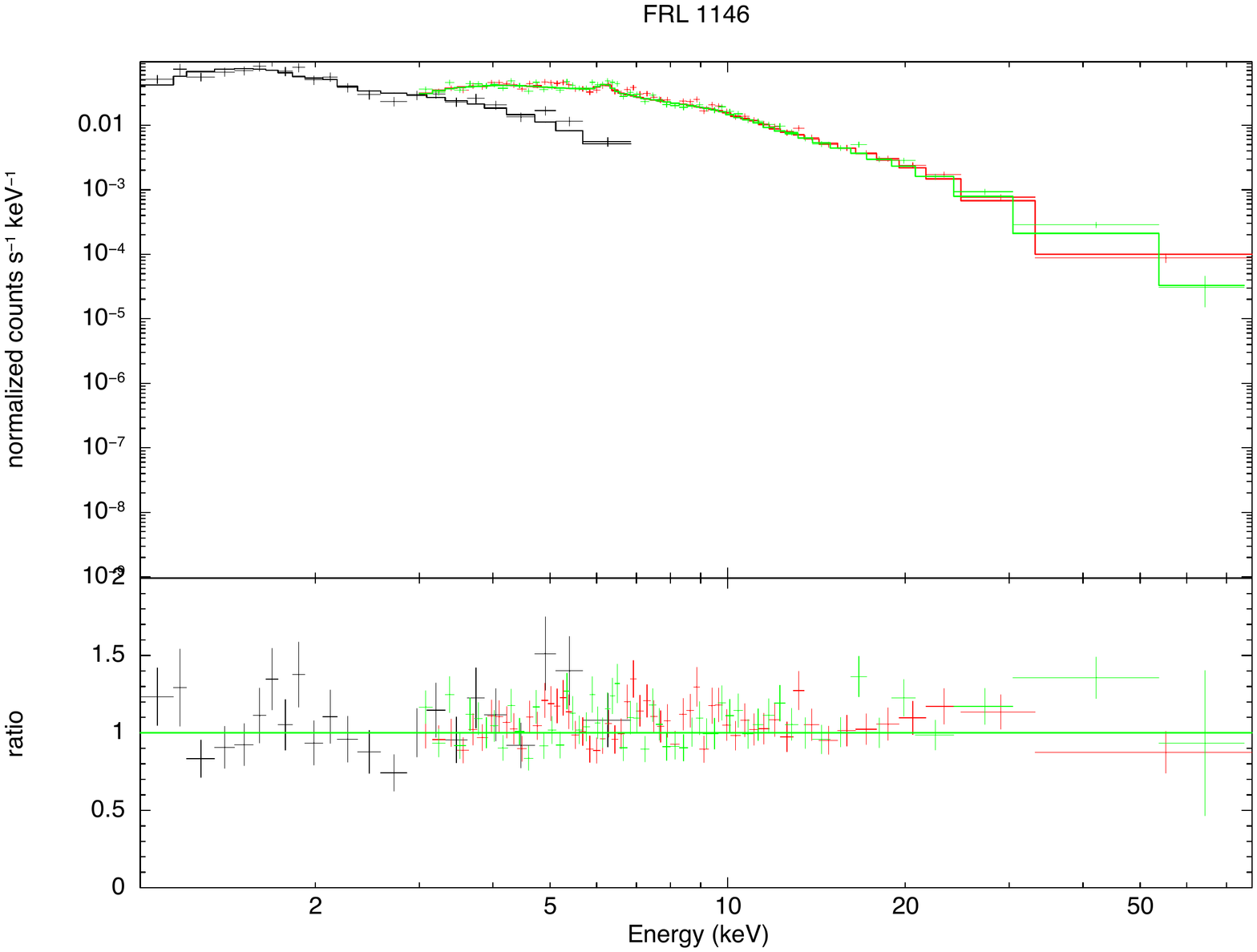}\\
\vspace{0.5cm}
\includegraphics[scale=0.25,angle=-90]{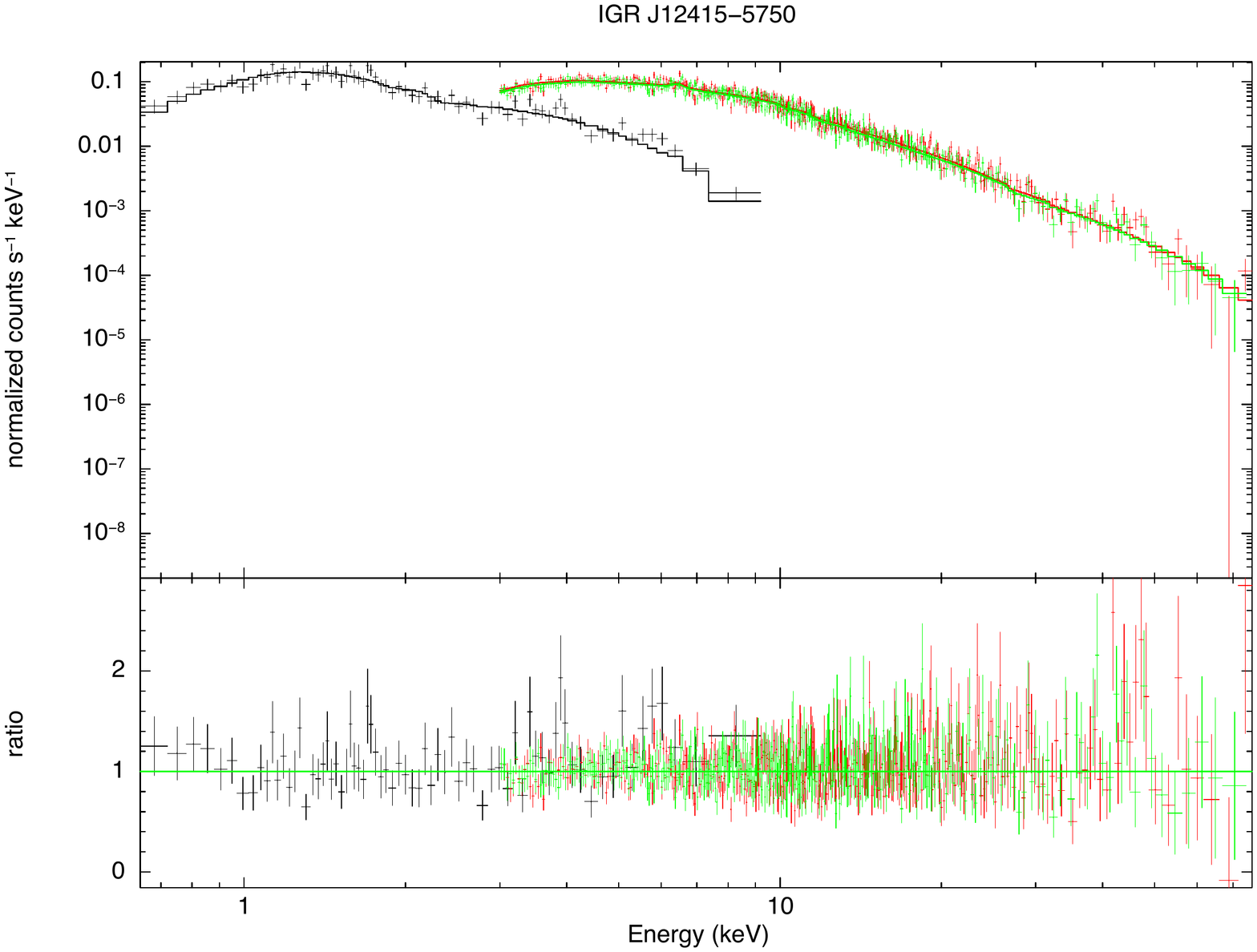}
\hspace{0.4cm}
\includegraphics[scale=0.25,angle=-90]{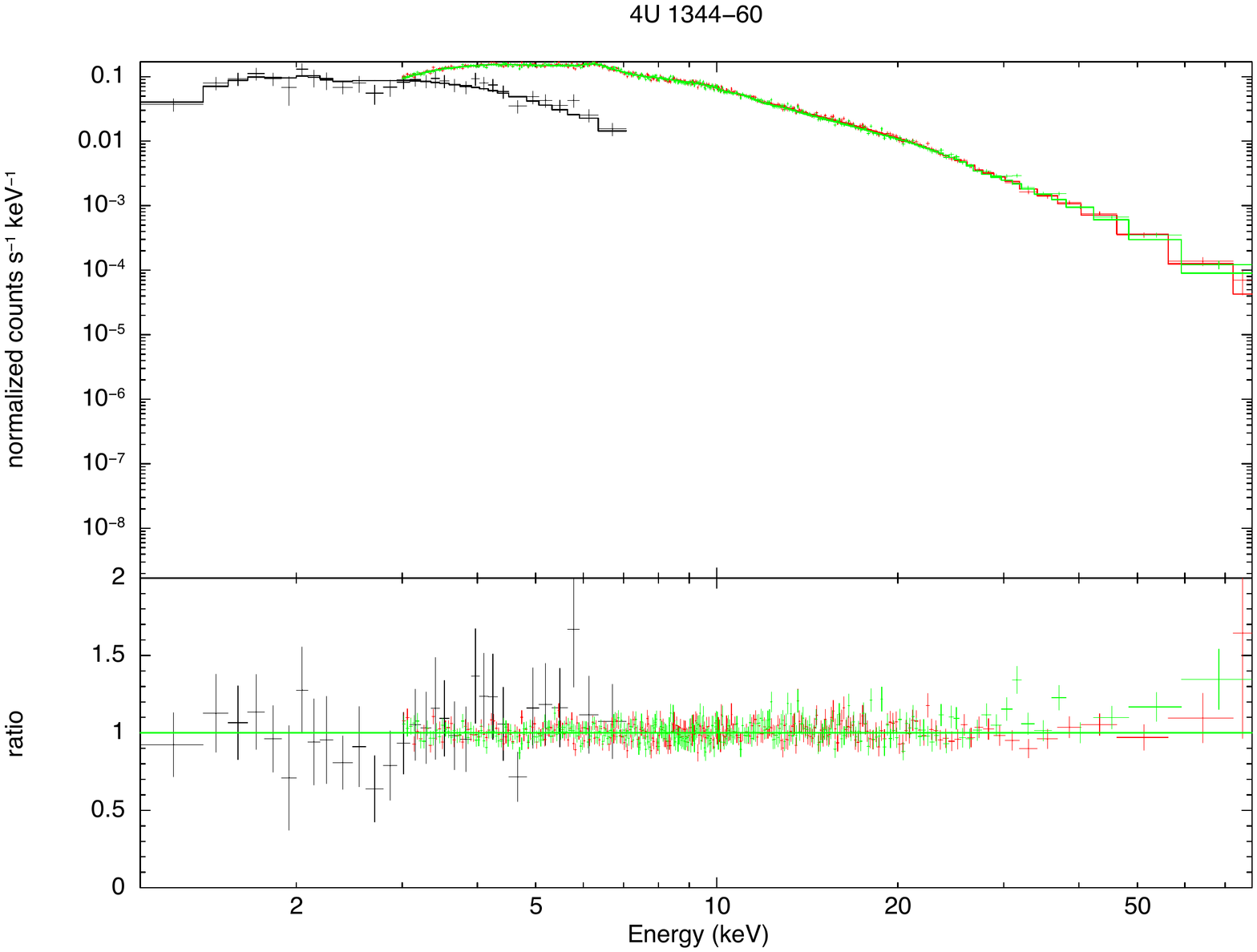}\\
\vspace{0.5cm}
\includegraphics[scale=0.25,angle=-90]{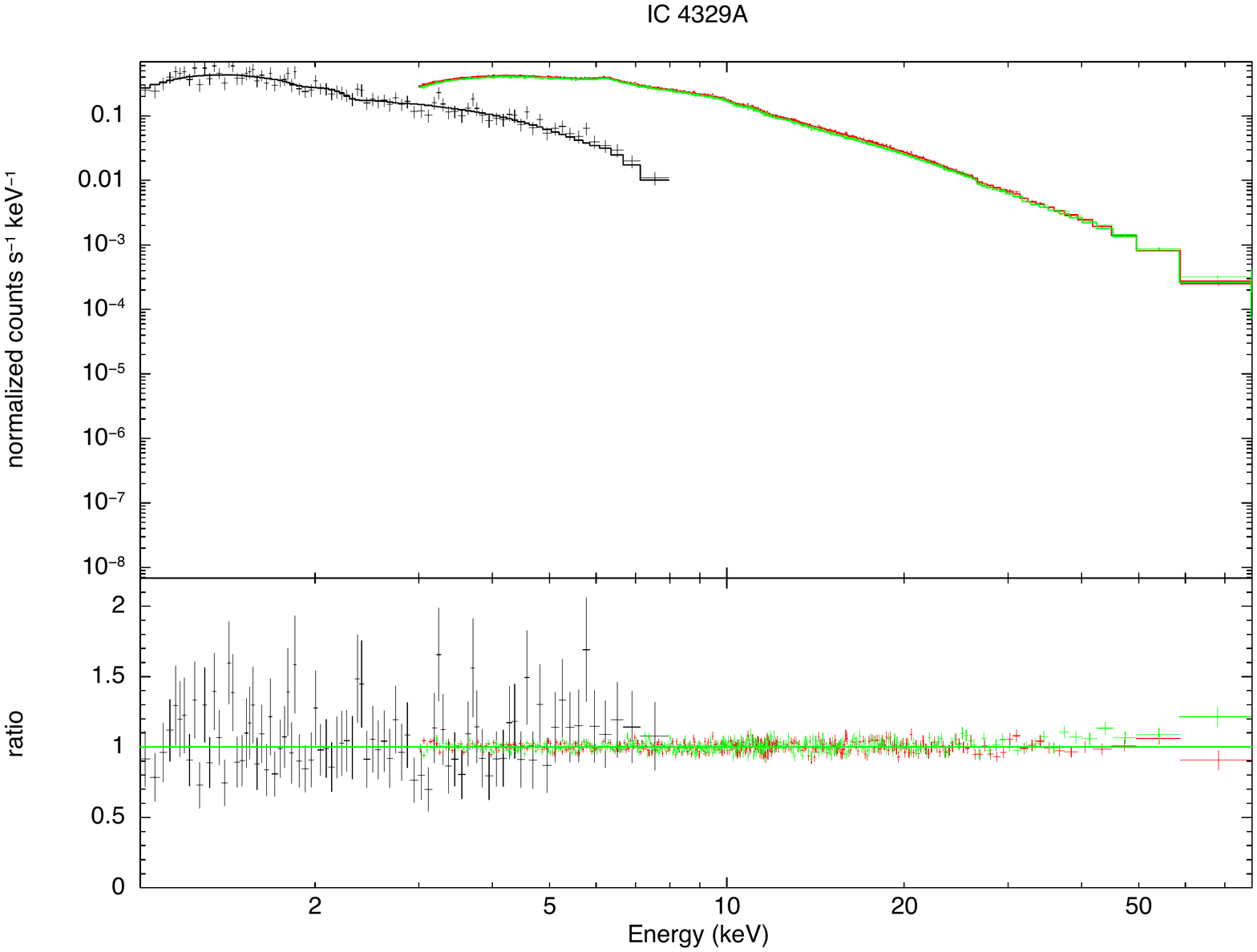}
\hspace{0.4cm}
\includegraphics[scale=0.25,angle=-90]{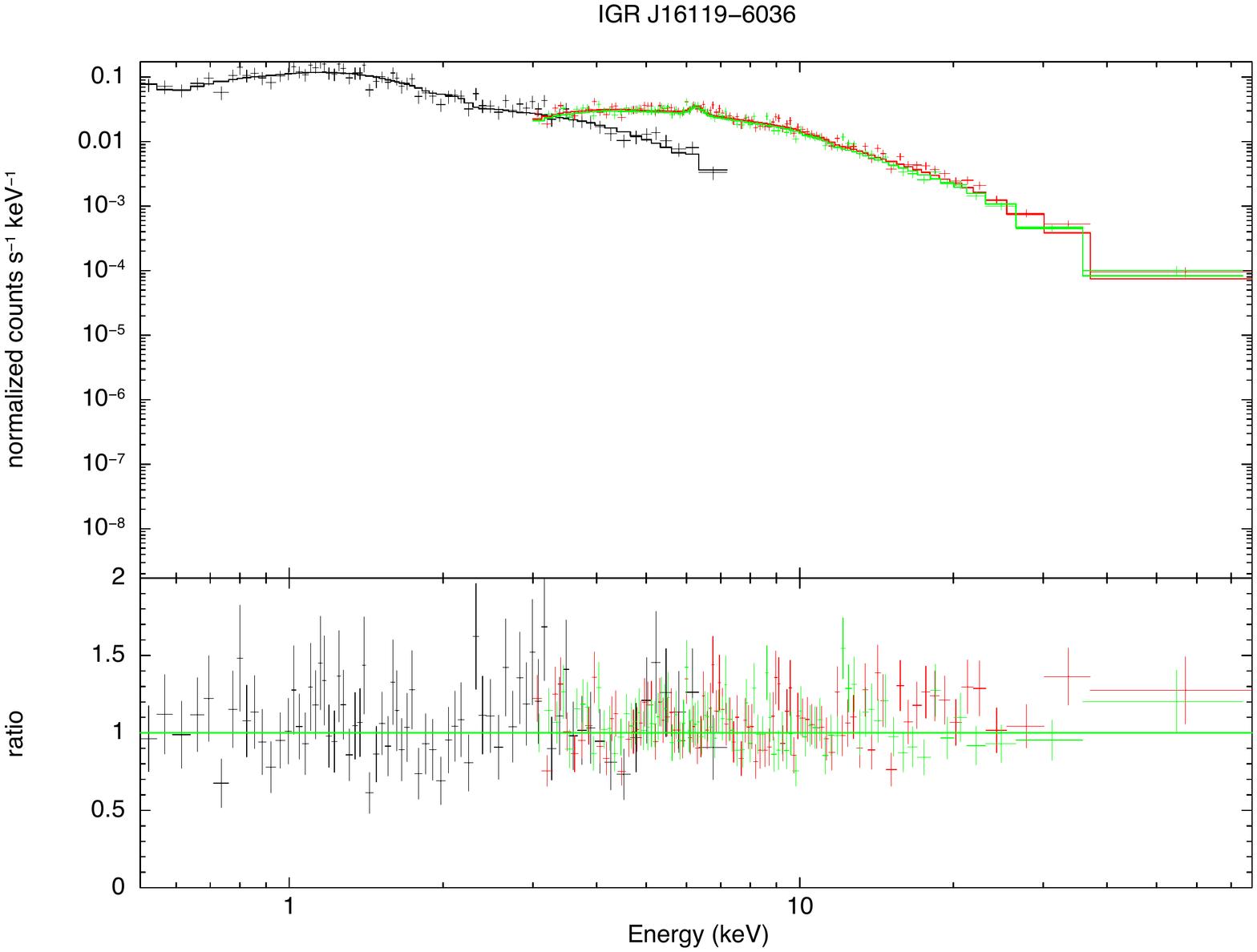}\\
\caption{{\small Data and folded model (upper panels) and model-to-data ratios (lower panel)
for QSO 0241+62, MCG+08-11-011, Mrk 6, FRL 1146, IGR J12415-5750, 4U 1344-60, IC 4329A
and IGR J16119-6036. Fit results are reported in Table 2.}}
\label{ic_16119}
\end{figure*}
\end{small}

\begin{small}
\begin{figure*}
\centering
\includegraphics[scale=0.25,angle=-90]{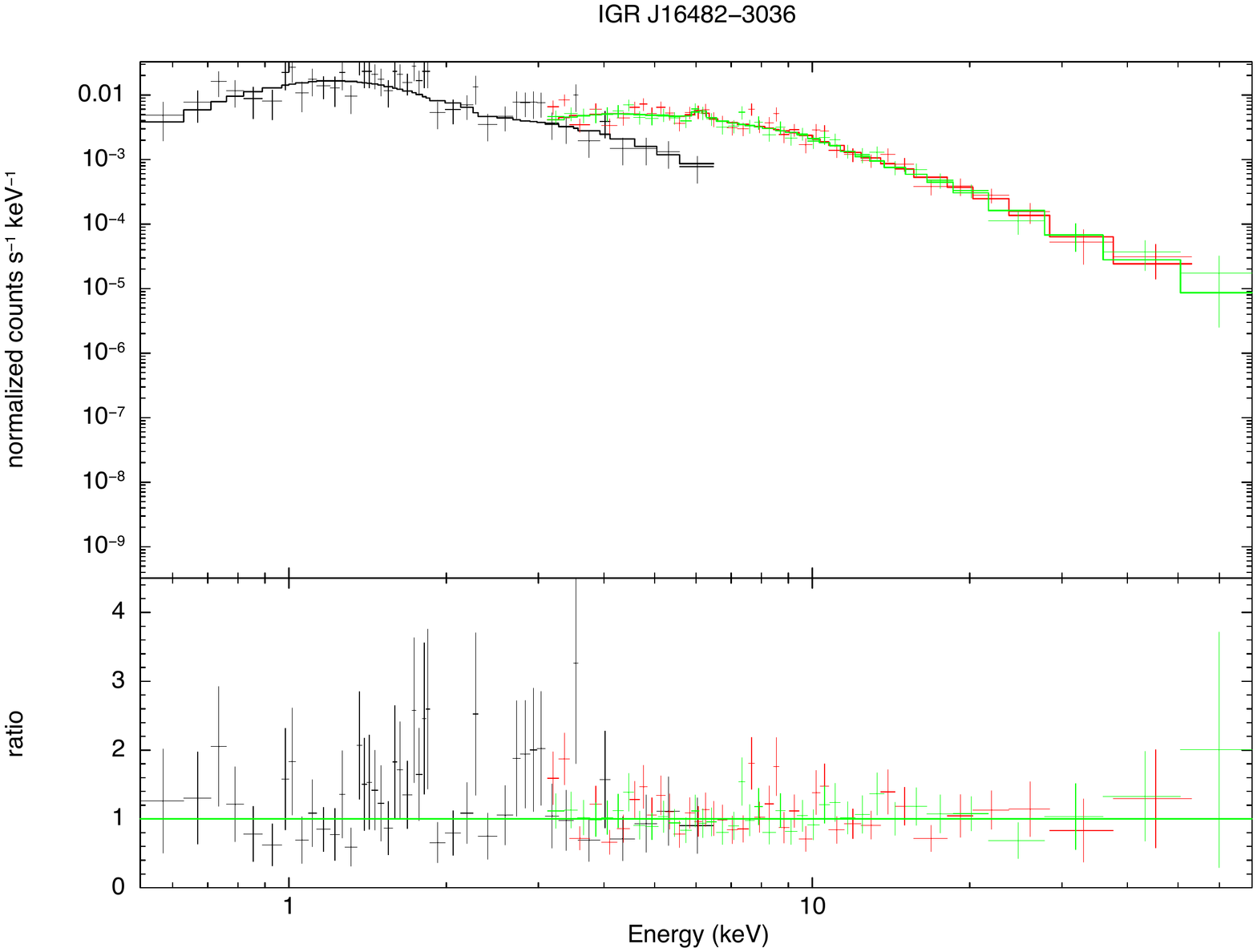}
\hspace{0.4cm}
\includegraphics[scale=0.25,angle=-90]{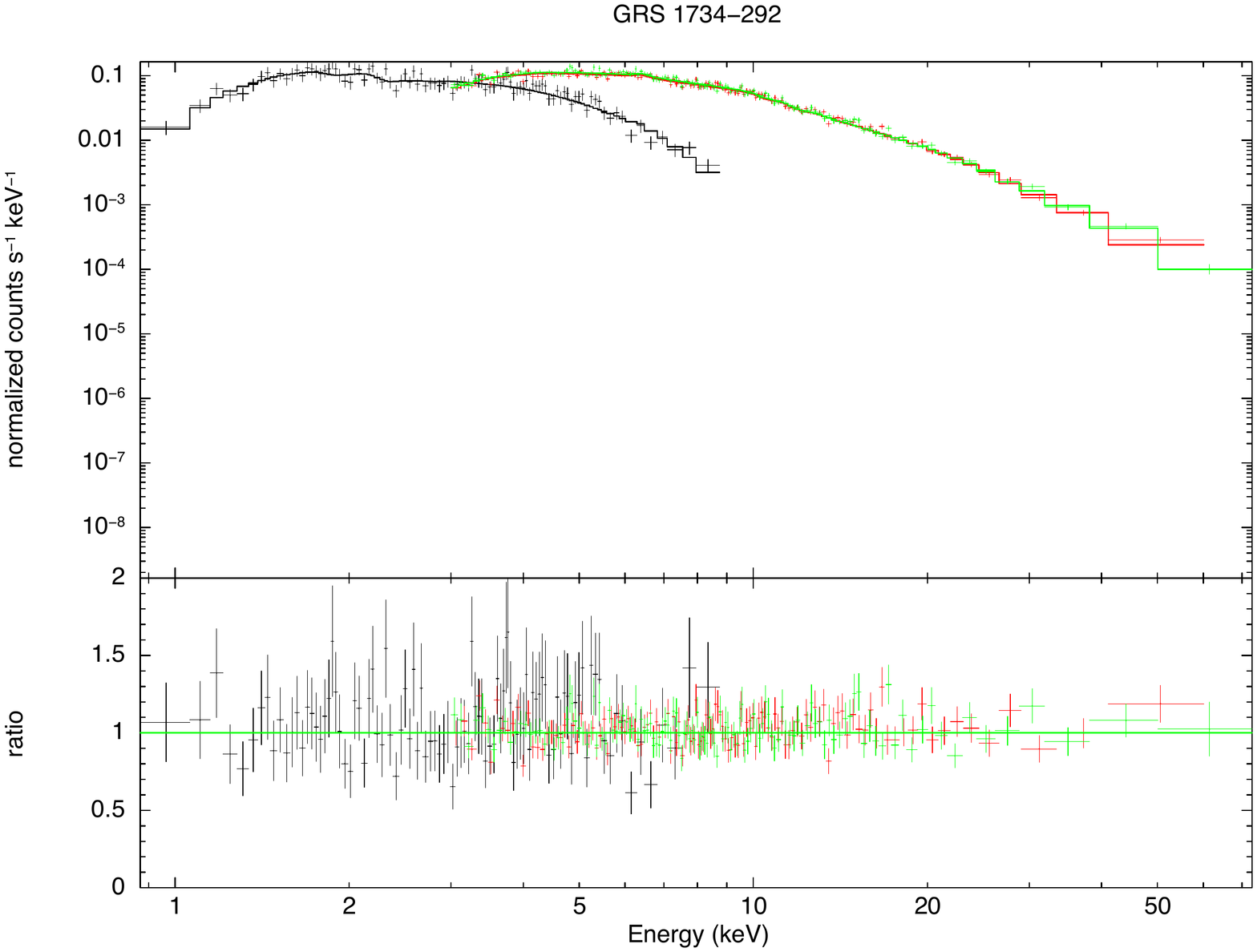}\\
\vspace{0.5cm}
\includegraphics[scale=0.25,angle=-90]{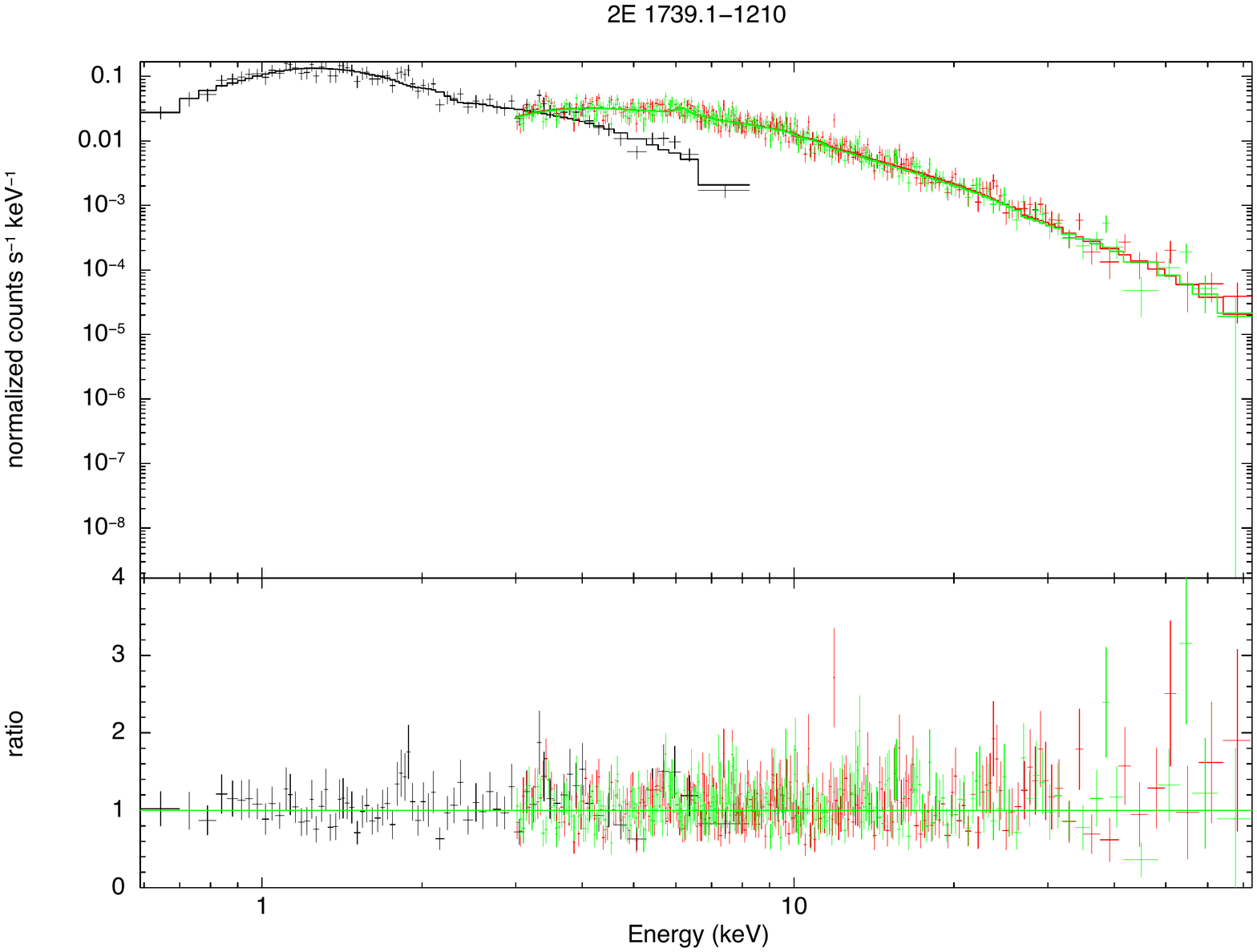}
\hspace{0.4cm}
\includegraphics[scale=0.25,angle=-90]{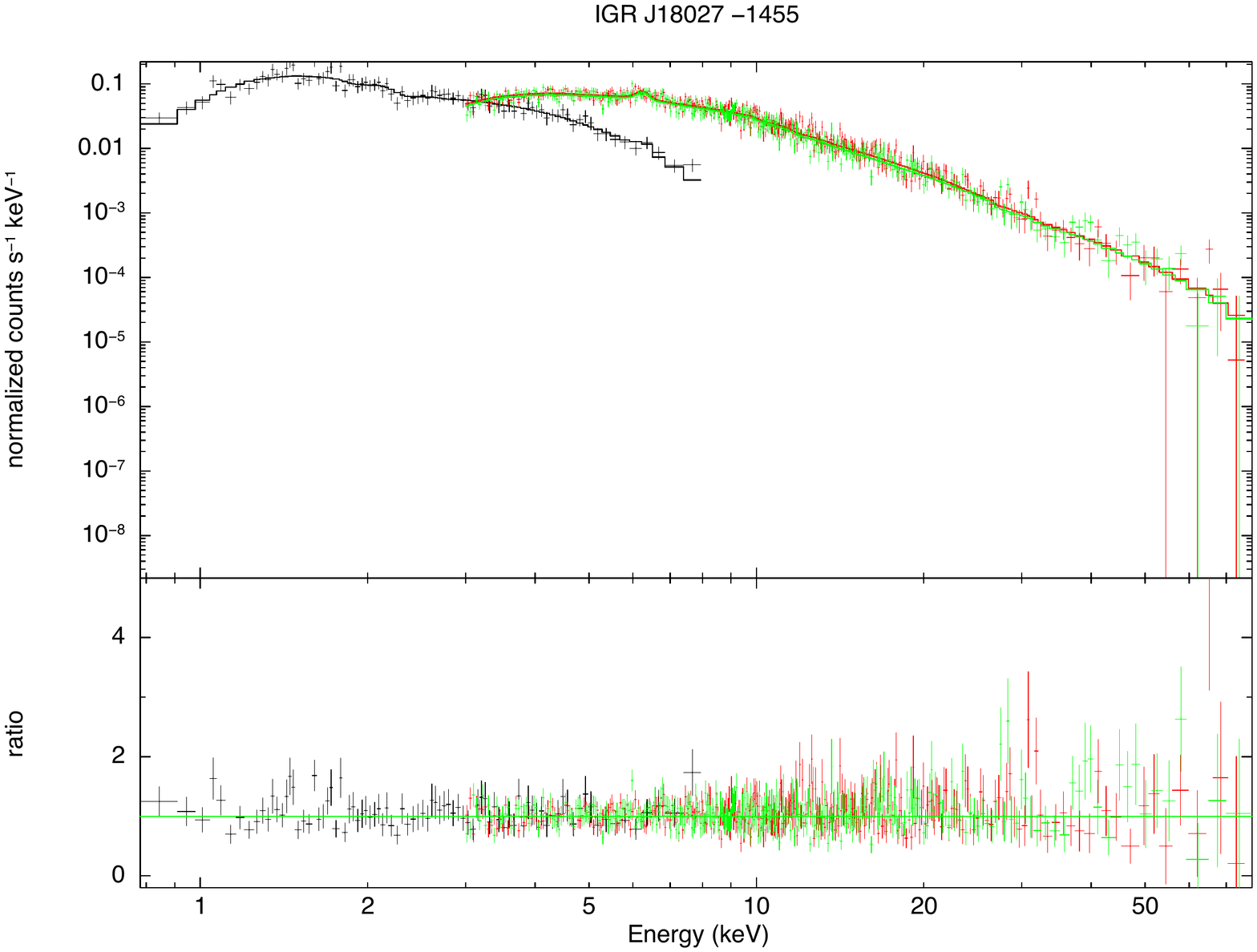}\\
\vspace{0.5cm}
\includegraphics[scale=0.25,angle=-90]{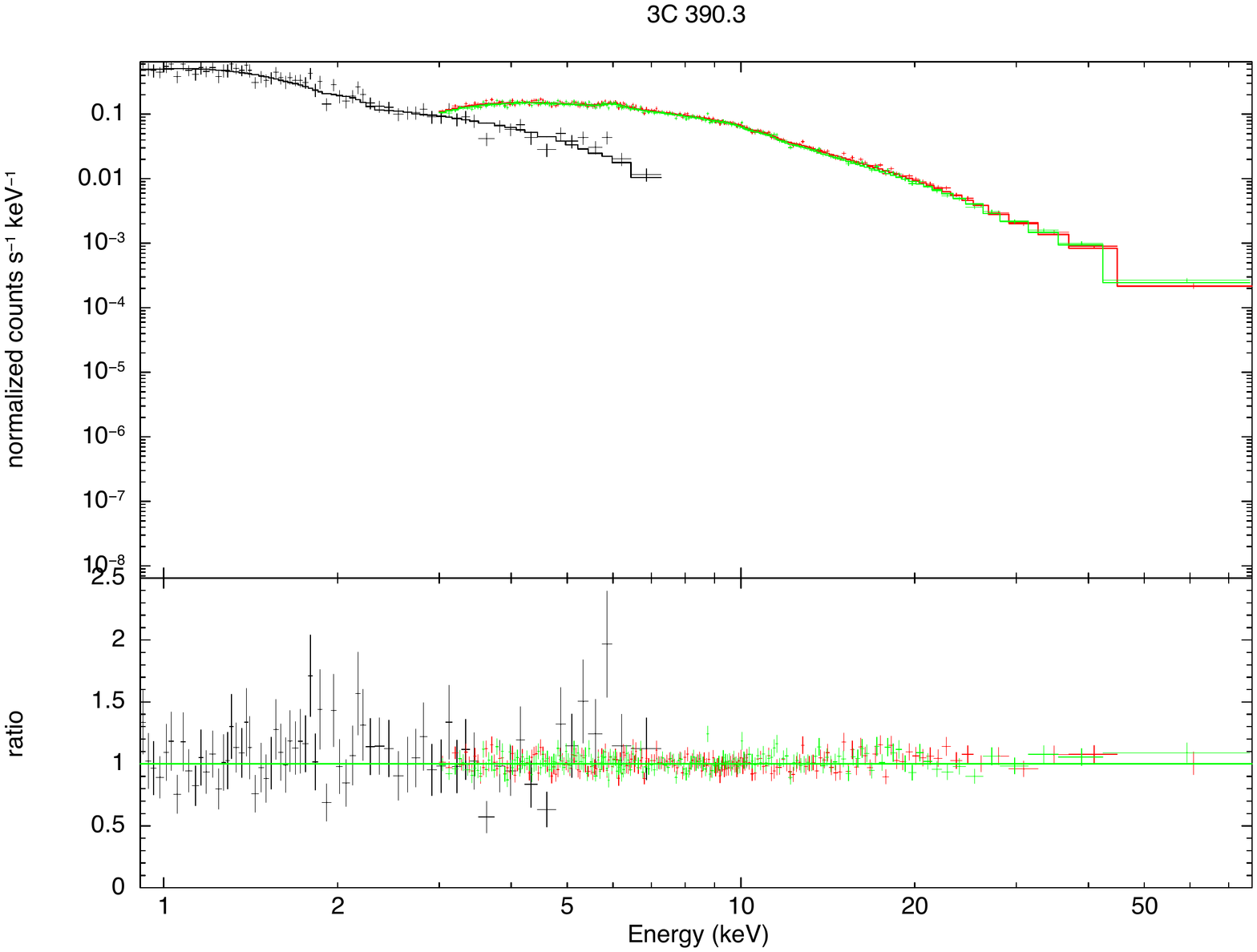}
\hspace{0.4cm}
\includegraphics[scale=0.25,angle=-90]{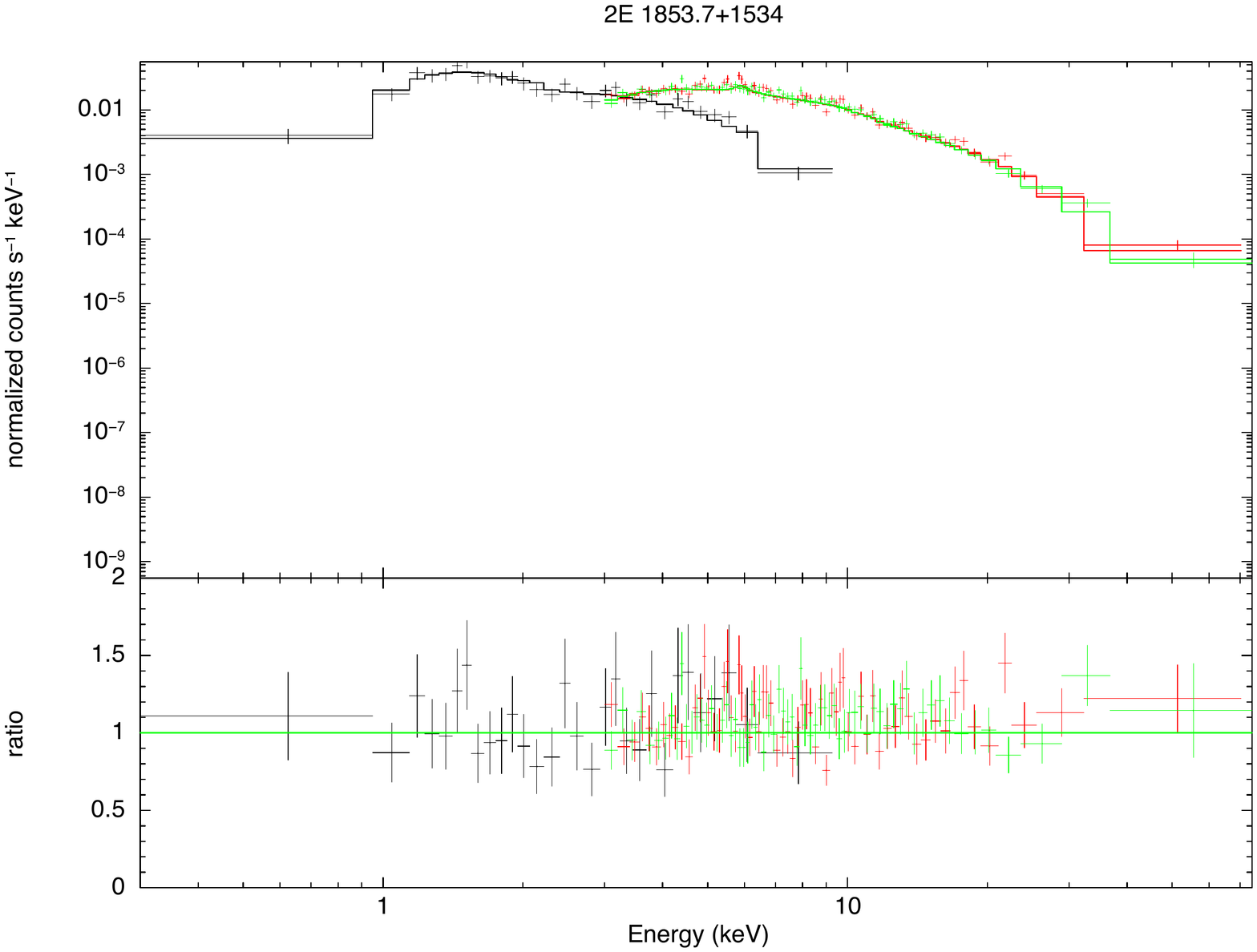}\\
\vspace{0.5cm}
\includegraphics[scale=0.25,angle=-90]{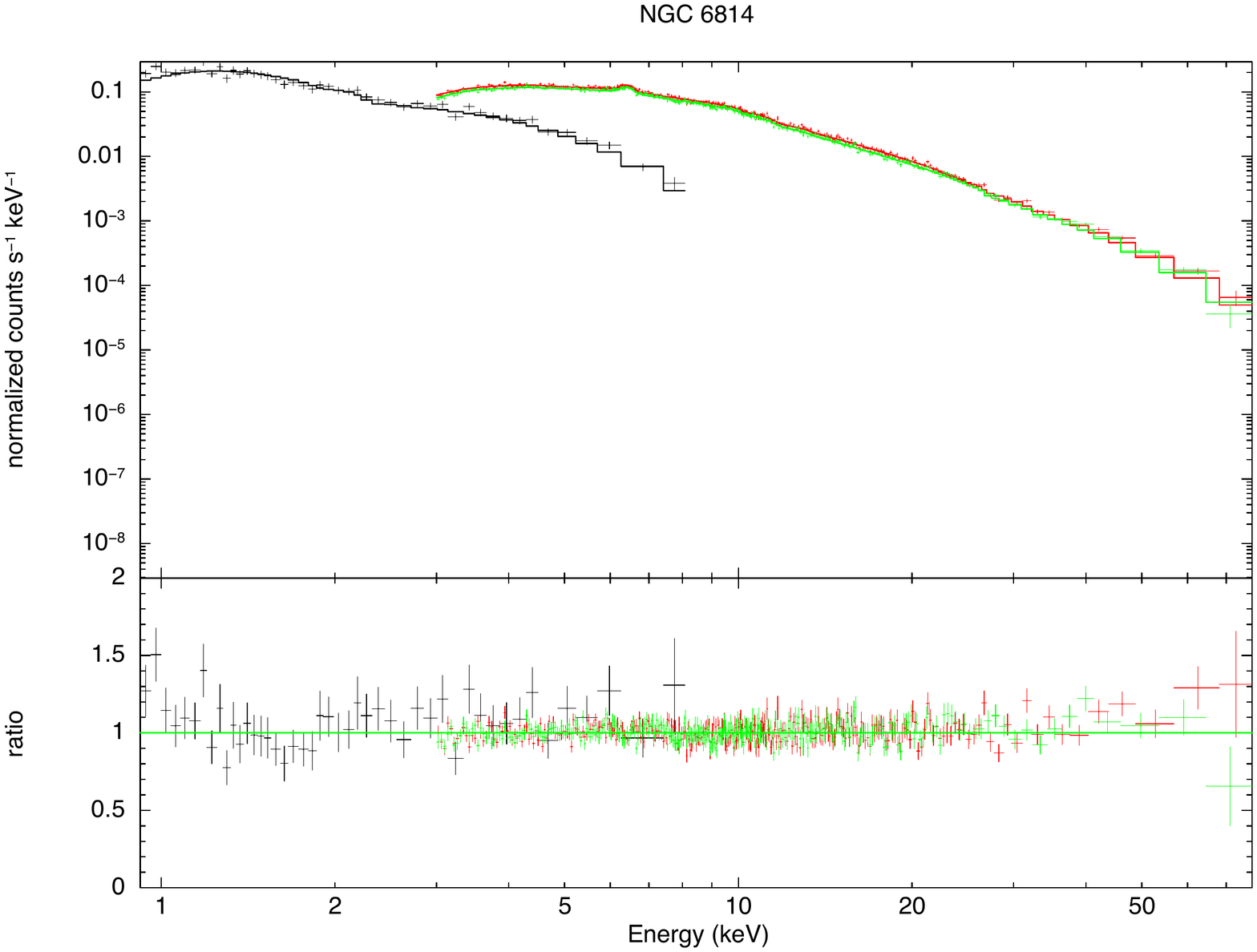}
\hspace{0.4cm}
\includegraphics[scale=0.25,angle=-90]{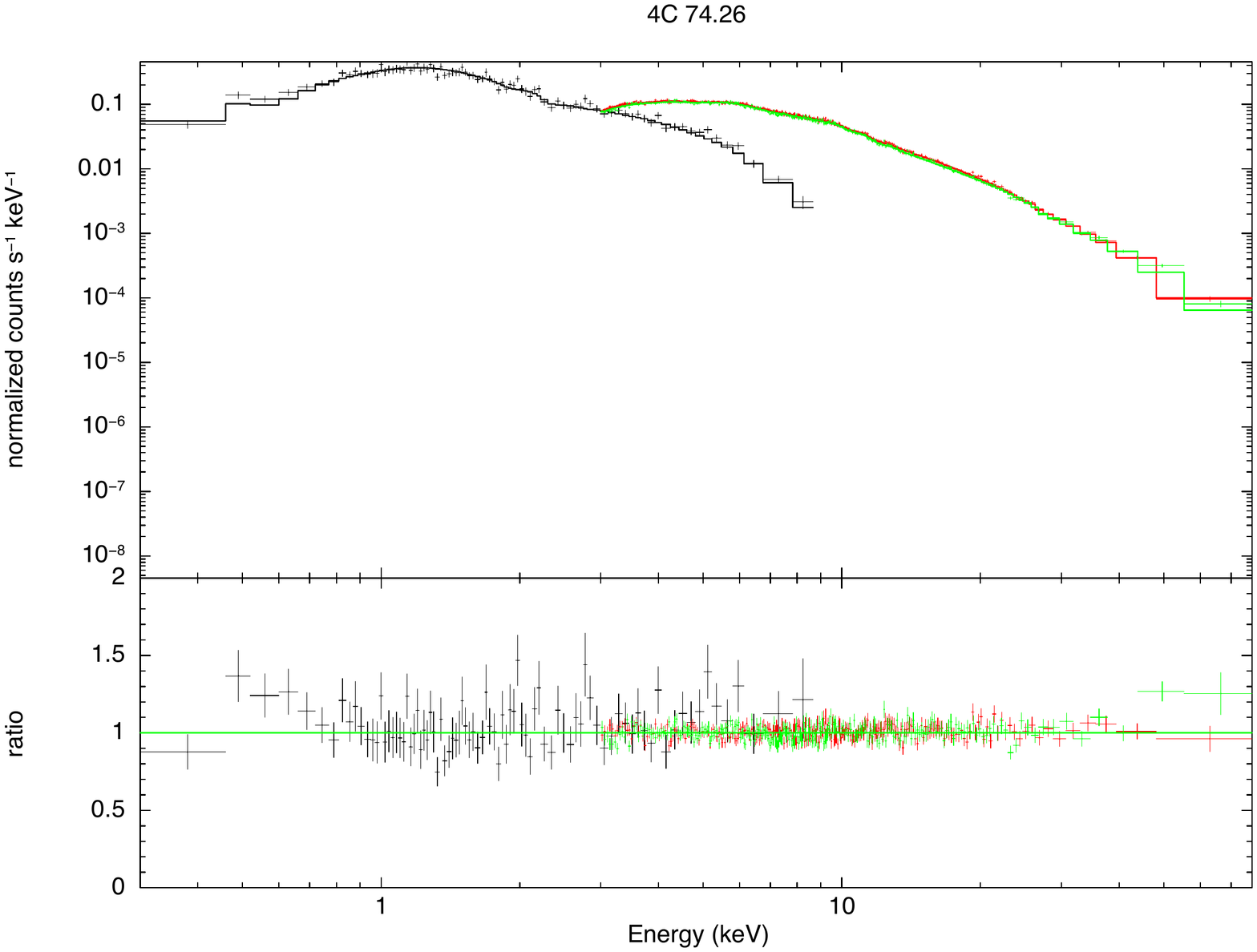}\\
\caption{{\small Data and folded model (upper panels) and model-to-data ratios (lower panel)
for IGR J16482-3036, GRS 1734-292,2E 1739.1-1210, IGR J18027-1455, 3C 390.3, 2E 1853.7+1534,
NGC 6814 and 4C 74.26 (summed observations). Fit results are reported in Table 2.}}
\label{6814_4c}
\end{figure*}
\end{small}

\begin{small}
\begin{figure*}
\centering
\includegraphics[scale=0.25,angle=-90]{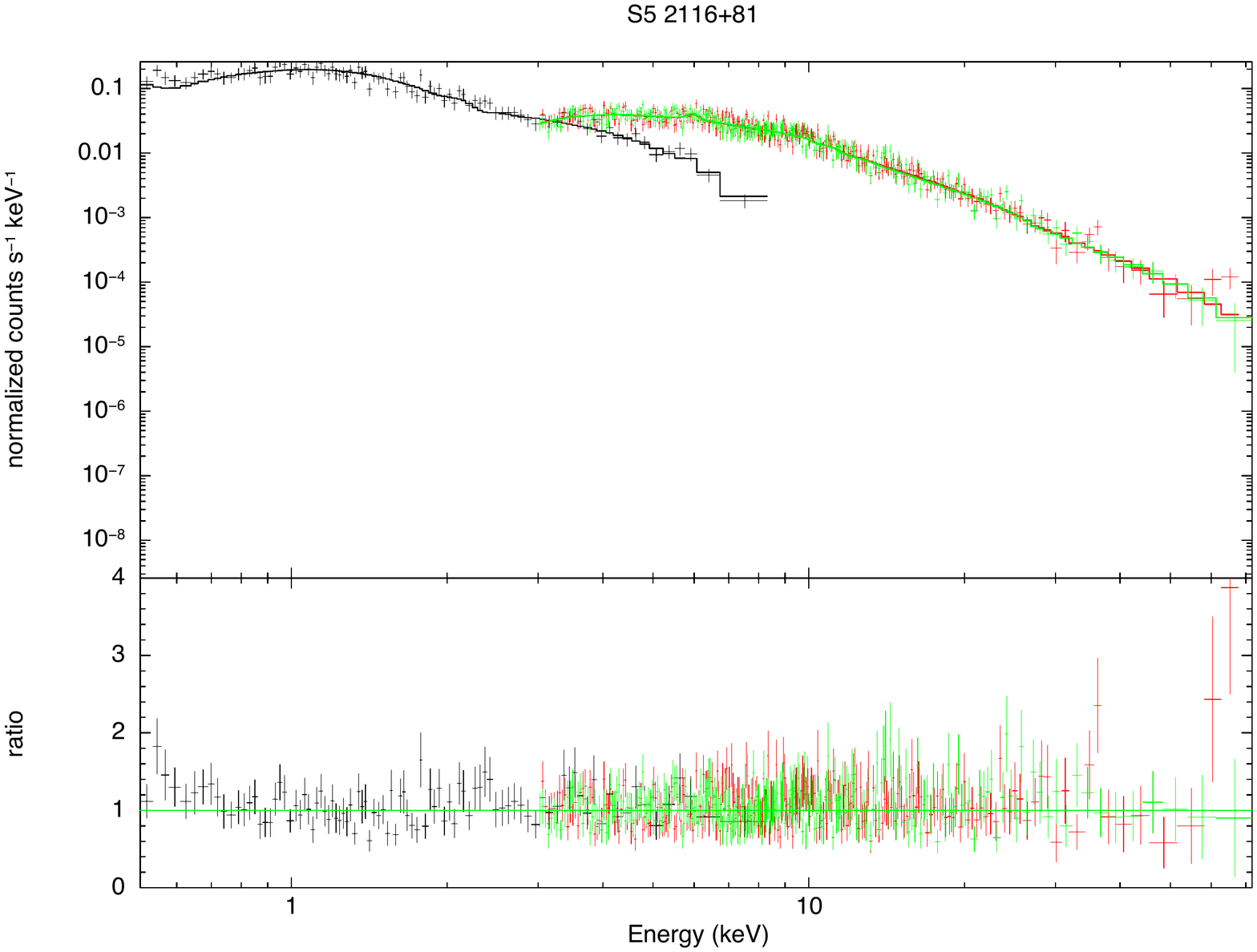}
\hspace{0.4cm}
\includegraphics[scale=0.25,angle=-90]{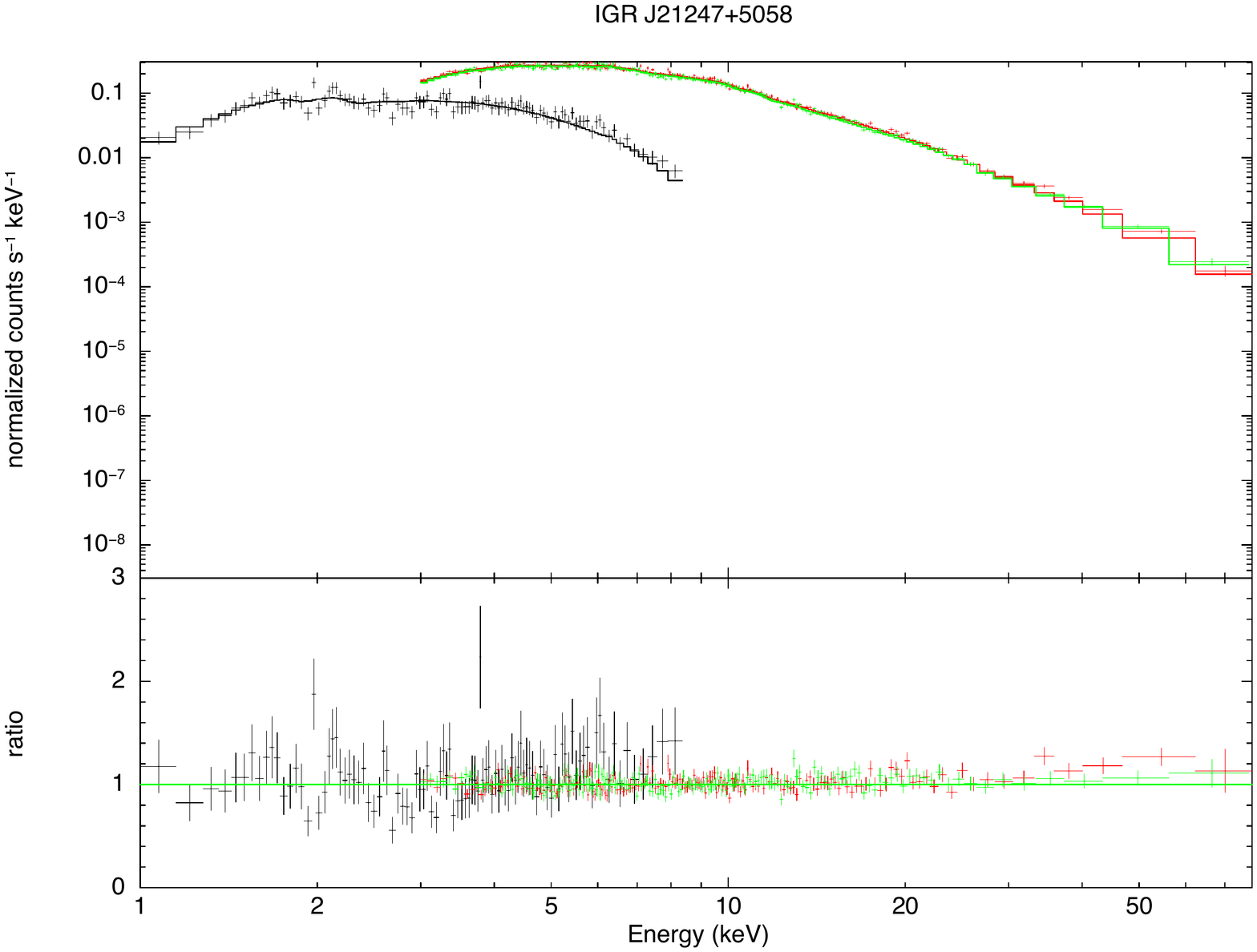}\\
\caption{{\small Data and folded model (upper panels) and model-to-data ratios (lower panel)
for S5 2116+81 and IGR J21247+5058. Fit results are reported in Table 2.}}
\label{2116_21247}
\end{figure*}
\end{small}

\begin{small}
\begin{figure*}
\centering
\includegraphics[scale=0.25,angle=-90]{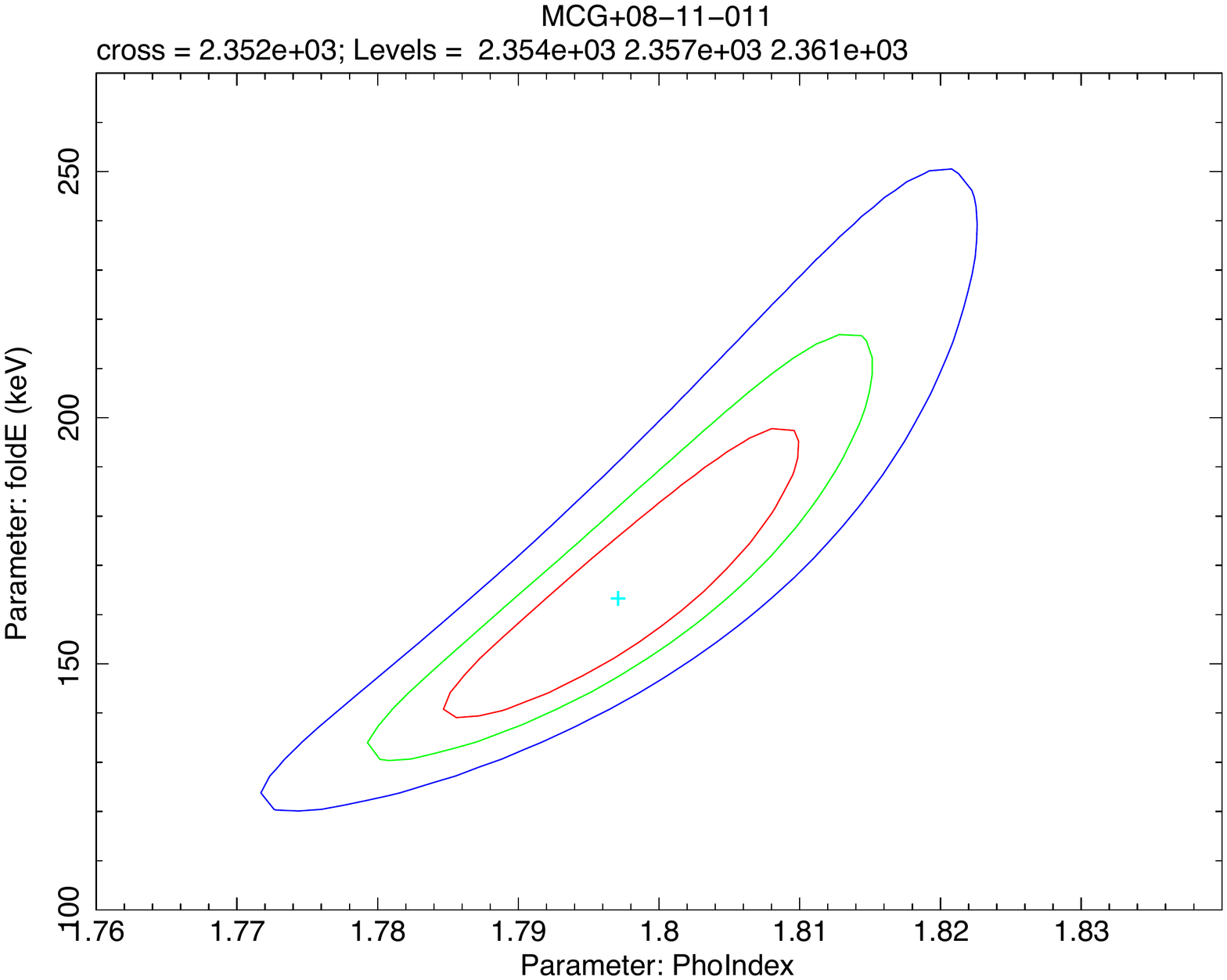}
\hspace{0.4cm}
\includegraphics[scale=0.25,angle=-90]{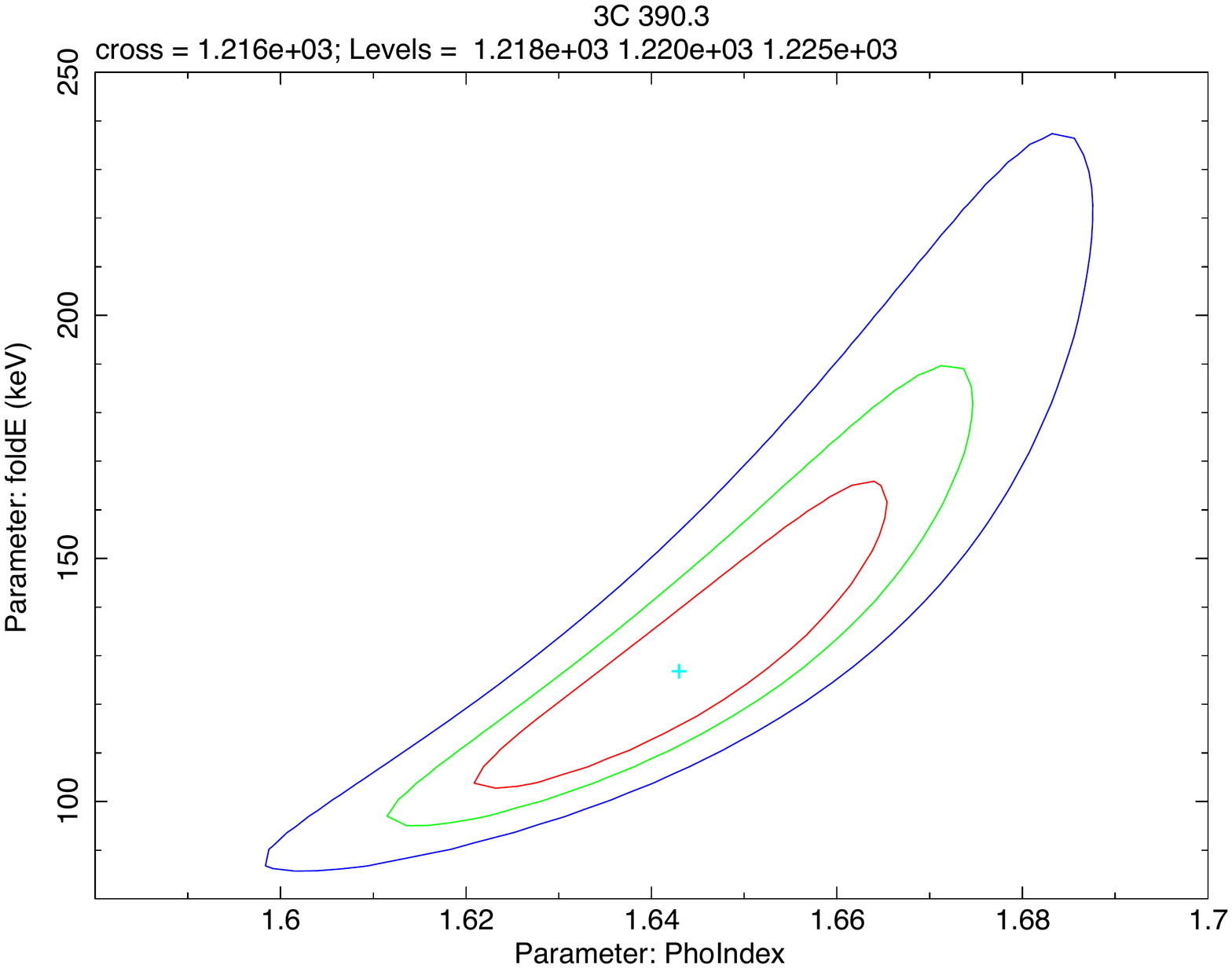}\\
\vspace{0.5cm}
\includegraphics[scale=0.25,angle=-90]{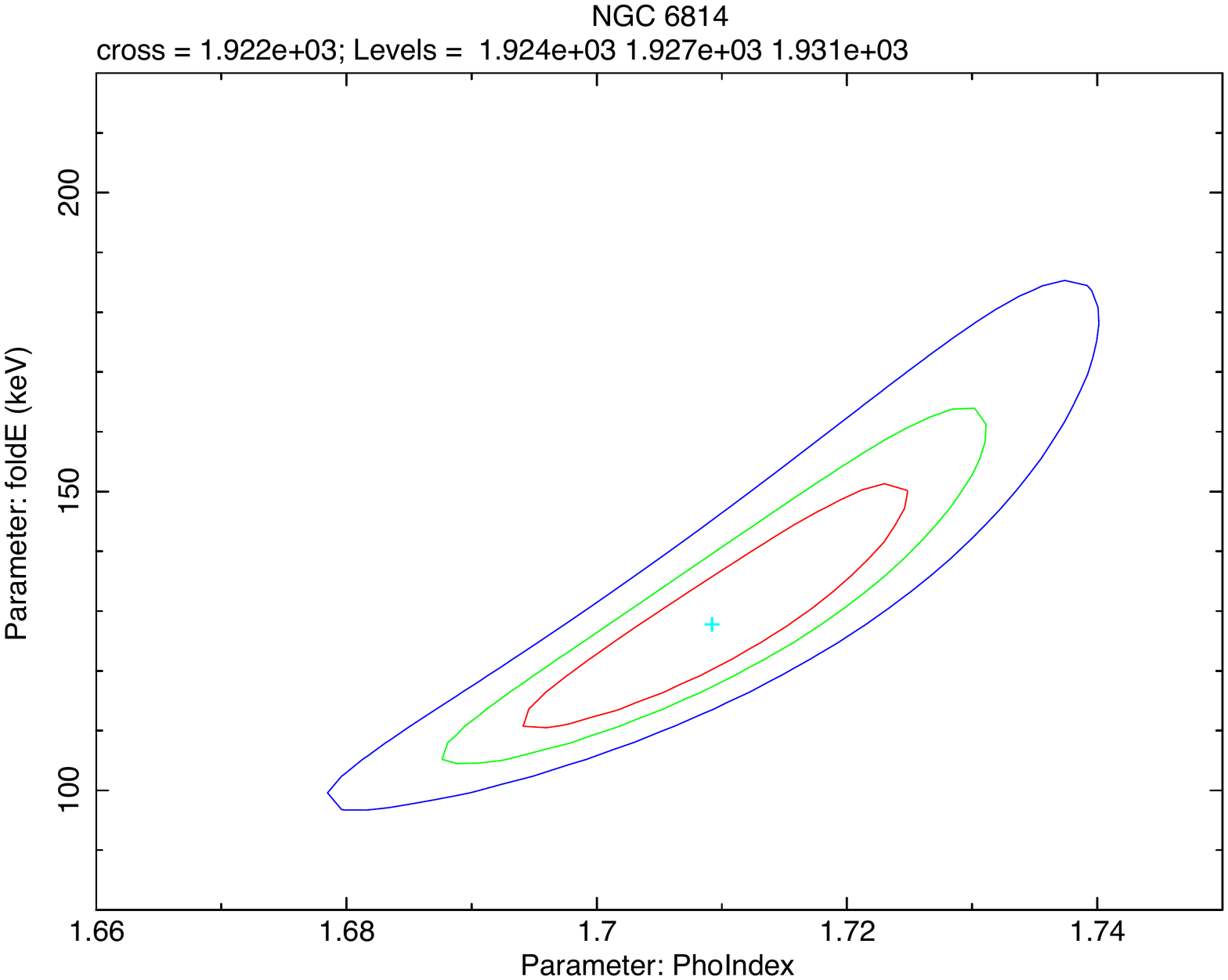}\\
\caption{{\small Confidence contour plots of the high energy cut-off vs. the photon
index for MCG+08-11-011 (upper left panel), 3C 390.3 (upper right panel) and NGC 6814 (lower panel).}}
\label{cont}
\end{figure*}
\end{small}




\bsp	
\label{lastpage}
\end{document}